\documentclass[lettersize,journal]{IEEEtran}
\usepackage{amsmath}
\usepackage{graphicx}
\usepackage{amsfonts}
\usepackage{subcaption}
\usepackage{amsthm}
\usepackage{amssymb}
\newtheorem{theorem}{Theorem}

\begin{document}

\title{Optimized Mirabbasi-Martin Pulse Shaping Filter for Higher-Order Modulation in Zak-OTFS}
\author{Naveed Bin Nazir and Ananthanarayanan Chockalingam \\
 Department of ECE, Indian Institute of Science, Bangalore 560012, India}
\maketitle

\begin{abstract}
The performance of Zak transform based orthogonal time frequency space (Zak-OTFS) modulation depends critically on the delay-Doppler (DD) domain pulse shaping filter, which must support both accurate channel estimation and reliable data detection. Filters reported in the Zak-OTFS literature (e.g., sinc, Gaussian, Gaussian-sinc (GS) filters) fall short in bit error rate (BER) performance for higher-order modulation due to sensitivity to model-free channel estimation errors and pilot-data interference. Motivated by this, in this paper, we introduce a separable two-dimensional Mirabbasi-Martin (MM) filter for Zak-OTFS and propose a filter coefficient optimization framework that minimizes the aggregate sidelobe energy while preserving the compact structure of the MM prototype. By trading exact Nyquist nulls for high localization, the optimized MM filter exhibits no sidelobes with the narrowest $-40$ dB DD domain pulsewidth compared to sinc, Gaussian, and GS pulses. This compactness of the optimized filter tightly confines the pilot energy within the pilot region of the embedded frame, ensuring that the full response is captured for model-free channel estimation while significantly reducing pilot-data interference. This yields both an improved estimation performance and a significantly improved BER performance for higher-order QAM. For the proposed filter, we  
derive closed-form expressions for the effective DD domain channel and the covariance of the receive-filtered noise, which drastically reduce simulation run times. Our simulation results demonstrate that the optimized MM filter achieves significantly better uncoded/coded BER performance for higher-order QAM compared to that of sinc, Gaussian, and GS filters. For example, for 64-QAM with $M=32, N=48$, rate-$1/2$ convolutional code, and TDL-A channel model with fractional DDs, the optimized MM filter achieves a coded BER of $3\times10^{-4}$ at an $E_\mathrm{b}/N_0$ of 40 dB, while the sinc, Gaussian, and GS filters exhibit error floors above $10^{-2}$.
\end{abstract}
\vspace{1mm}

\section{Introduction}
Orthogonal time frequency space (OTFS) modulation is a delay-Doppler (DD) domain modulation that has attracted considerable interest for wireless communication over doubly-dispersive channels. OTFS has evolved from its early multicarrier (MC) formulation \cite{mc_otfs1}-\cite{mc_otfs4} to the more recent Zak-transform-based OTFS (Zak-OTFS) framework \cite{zak_otfs1},\cite{zak_otfs2}. Zak-OTFS provides a direct transformation between the DD domain and time domain through the inverse Zak transform at the transmitter and vice-versa using the Zak transform at the receiver. This provides a rigorous mathematical framework for describing the DD domain waveform and its input-output (I/O) relation, while offering robustness over a wide range of delay and Doppler spreads \cite{zak_otfs3}-\cite{zak_otfs8}. Zak-OTFS provides a predictable and non-fading effective DD domain channel representation, which facilitates simple, model-free I/O relation estimation. A key component of the Zak-OTFS transceiver is the DD domain pulse shaping filter, which controls the time and bandwidth containment of the transmitted waveform and shapes the effective DD domain channel response given by the cascade of the transmit filter, physical channel, and receive filter.

In Zak-OTFS, a good DD pulse shaping filter should satisfy two main requirements: $i)$ accurate I/O relation estimation and $ii)$ reliable data detection. Model-free estimation using embedded pilot frames benefits from strong pulse localization, as concentrated pulse energy and fast sidelobe decay confine the pilot response within the allotted pilot region, preventing channel response truncation and limiting pilot-data interference. Data detection, on the other hand, benefits from pulse orthogonality, where nulls at non-zero DD grid-points avoid inter-symbol interference (ISI). The commonly considered sinc and Gaussian pulses represent two different ends of this orthogonality-localization tradeoff. The sinc pulse satisfies the Nyquist criterion and has nulls at non-zero DD grid-points, but its slowly decaying sidelobes result in poor DD localization. In contrast, the Gaussian pulse provides good DD localization with no sidelobes, but lacks nulls at non-zero DD grid-points and therefore introduces ISI \cite{zak_otfs5}. The Gaussian-sinc (GS) pulse combines the favorable properties of Gaussian and sinc pulses without expanding the time or bandwidth support \cite{zak_otfs6}, though its effective pulse spread remains significantly broader than that of the Gaussian pulse. More recently, Hermite-basis pulse design has provided greater flexibility in controlling localization and orthogonality through basis-function and coefficient optimization \cite{zak_otfs7}. The GS pulse has been shown to achieve better bit error rate (BER) performance with model-free estimation compared to sinc and Gaussian pulses \cite{zak_otfs6}, and the optimized Hermite pulse has been reported to achieve BER performance comparable to that of the GS pulse \cite{zak_otfs7}.

Reported BER performance evaluations for different DD pulse shapes in Zak-OTFS have focused primarily on lower-order modulation (e.g., BPSK/4-QAM/8-QAM). However, there has not been much focus on higher-order modulation (e.g., 64-QAM$-$256-QAM), which is of interest. Our performance evaluations of existing DD pulses for higher-order QAM revealed poor BER performance with floors at high error rates. This motivated us to explore other pulse shapes that can perform well for higher-order modulation. The work reported in this paper is a positive outcome of this exploration. Towards achieving this, we found that prioritizing strong DD localization and aggregate sidelobe suppression in the pulse design, while selecting the perturbation bound to retain reasonably low responses at non-zero DD grid-points (rather than demanding exact nulls), proves to be more effective. Guided by the above motivation and rationale, we explore pulse designs that can offer the desired localization-orthogonality tradeoff under the Mirabbasi-Martin filter design framework.

The Mirabbasi-Martin (MM) prototype filter, introduced in \cite{MM_1}, was originally developed for overlapped complex-modulated transmultiplexers to provide high stopband attenuation, fast sidelobe decay, and near-perfect reconstruction. The MM construction has subsequently been used in filter-bank multicarrier (FBMC) systems, including FBMC with offset quadrature amplitude modulation (FBMC/OQAM) and QAM-FBMC\footnote{A prominent realization is the prototype filter developed under the European PHYDYAS (Physical Layer for Dynamic Spectrum Access and Cognitive Radio) project, which corresponds to a specialized MM design with an overlap factor of $g=4$ optimized for frequency-domain stopband attenuation and near-perfect reconstruction \cite{MM_3}, \cite{MM_6}.}, owing to its favorable time-frequency localization, spectral confinement, and low sidelobe characteristics \cite{MM_2}-\cite{MM_6}. The MM structure has also been considered in practical waveform designs, e.g., for low peak-to-average power ratio (PAPR) preambles \cite{MM_7}. These properties, together with the structural flexibility offered by varying the overlap factor $g$ under strict time and frequency containment, motivate us to adapt the MM prototype into a separable two-dimensional DD pulse shaping filter with good delay and Doppler containment and tailor its coefficients for Zak-OTFS. The resulting optimized MM pulse exhibits no sidelobes and achieves the strongest DD localization among the considered benchmark pulses, enabling superior model-free channel estimation with negligible pilot-data interference, while keeping the response at non-zero DD grid-points low enough to deliver order-of-magnitude BER improvements for higher-order QAM. The key contributions in this paper are summarized as follows.
\begin{itemize}
    \item We develop a separable two-dimensional MM pulse in the DD domain for Zak-OTFS by extending the one-dimensional MM prototype to the delay and Doppler dimensions. We characterize the resulting pulse family across different overlap factors $g$ in terms of DD domain compactness, sidelobe behavior, and DD grid-point suppression, showing that lower overlap factors (specifically $g=3$) provide the most compact footprint while maintaining DD grid-point suppression comparable to higher overlap factors.

    \item We propose a coefficient optimization framework for the MM pulse family that minimizes aggregate sidelobe energy while constraining deviation from the prototype filter. We specialize this framework for the $g=3$ pulse by prioritizing sidelobe minimization over exact DD grid-point nulls, yielding an optimized pulse with no sidelobes and the narrowest $-40$ dB pulsewidth among the benchmark pulses (sinc, Gaussian, and GS pulses), while satisfying the $99\%$ energy-containment requirement.

    \item For the proposed MM pulse, we derive closed-form analytical expressions for the effective DD domain channel and the covariance of the receive-filtered DD domain noise. These closed-form expressions drastically reduce simulation run times (e.g., evaluating the effective DD-domain channel over $100$ channel realizations takes a simulation run time of $7.71$~s with closed-form expressions, whereas the same takes $253.78$~s with numerical integration). 

    \item We carry out extensive simulations to evaluate the performance of the proposed MM filter against benchmark sinc, Gaussian, and GS filters in the Vehicular-A channels with fractional DD shifts, embedded pilot frames, model-free channel estimation, and minimum mean square error (MMSE) detection. Our results show that, by confining pilot energy and substantially reducing pilot-data interference, the proposed MM filter achieves superior channel estimation and BER performance for higher-order QAM. For example, for 64-QAM with $M=32, N=48$, rate-$1/2$ convolutional code, and TDL-A channel model with fractional DDs \cite{tdla}, the optimized MM filter achieves a coded BER of $3 \times 10^{-4}$ at an $E_\mathrm{b}/N_0$ of 40 dB, while the sinc, Gaussian, and GS filters exhibit error floors above $10^{-2}$.
\end{itemize}
The rest of the paper is organized as follows. Section~\ref{sec:system} introduces the Zak-OTFS system model and its DD domain I/O relation. Section~\ref{sec:proposed} presents the proposed two-dimensional MM pulse, analyzes its DD domain characteristics, and develops the coefficient optimization procedure. Section~\ref{sec:closed_form} derives closed-form expressions for the effective DD domain channel and the covariance of the receive-filtered DD domain noise. Section~\ref{sec:results} presents simulation results and performance comparisons under different modulation orders and system configurations. 
Section~\ref{sec:conclusions} presents the conclusions. 

\section{Zak-OTFS System Model}
\label{sec:system}
\begin{figure}[t]
\includegraphics[width=8.5cm, height=4.25cm]{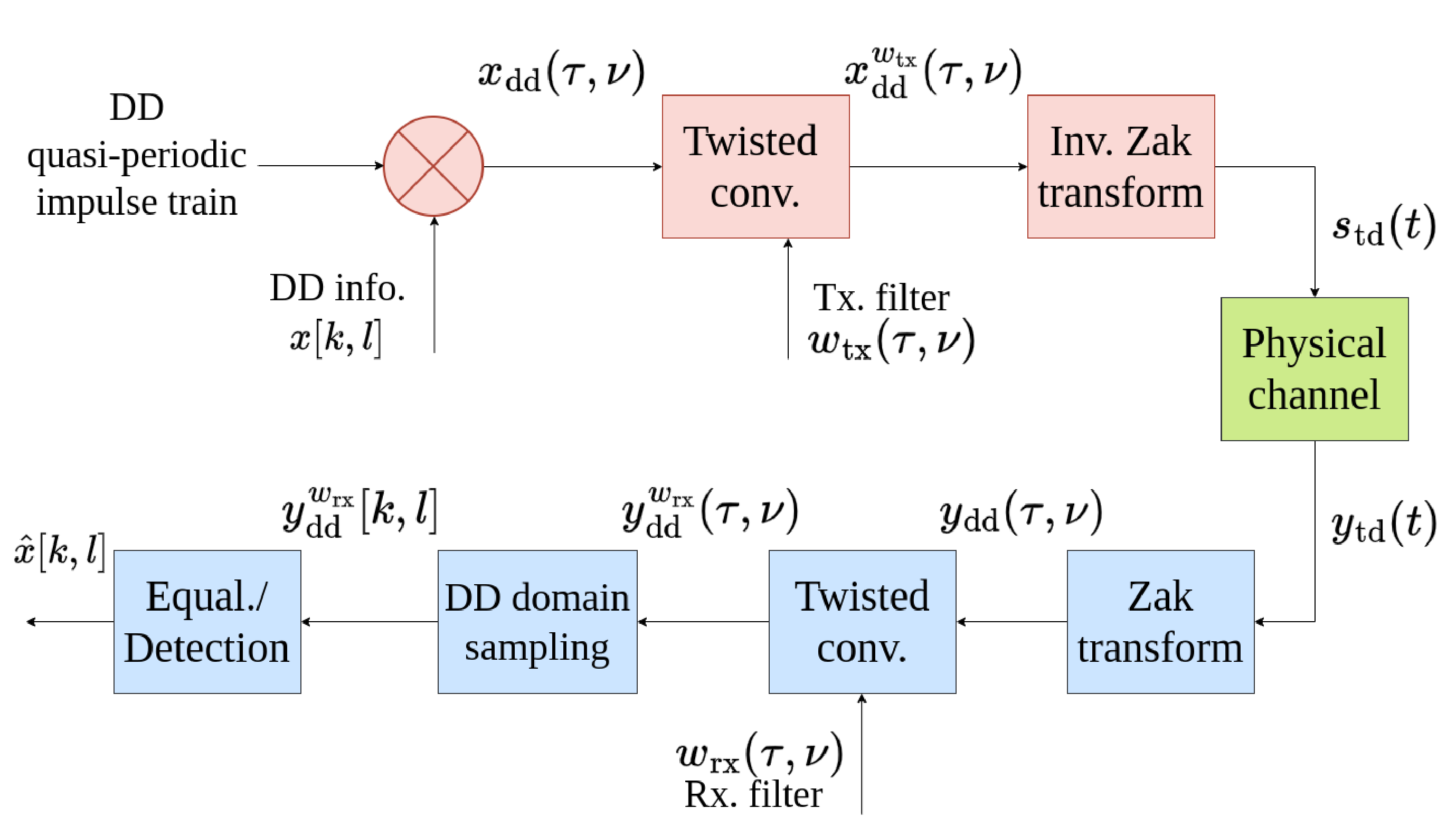}
\caption{Zak-OTFS transceiver block diagram.} 
\label{fig:zak_otfs_block_diagram}
\vspace{-4mm}
\end{figure}

The Zak-OTFS transceiver is shown in Fig.~\ref{fig:zak_otfs_block_diagram}. In Zak-OTFS, information symbols are multiplexed in the DD domain, converted into time domain, and transmitted. The basic information carrier is a DD domain pulse, which is a quasi-periodic function characterized by a delay period $\tau_{\mathrm{p}}$ and a Doppler period $\nu_{\mathrm{p}}$ satisfying $\tau_{\mathrm{p}} \nu_{\mathrm{p}} = 1$. The fundamental region in the DD domain is defined as $\mathcal{D}_0
=
\left\{(\tau,\nu):0\leq\tau<\tau_{\mathrm{p}},\;
0\leq\nu<\nu_{\mathrm{p}}\right\}$, where $\tau$ and $\nu$ denote the delay and Doppler variables, respectively. The fundamental region is discretized into $M$ bins along the delay dimension and $N$ bins along the Doppler dimension, giving the DD lattice $\Lambda_{\rm DD} = \left\{\left(k\Delta\tau, l\Delta\nu\right)\,\middle|\,k,l\in\mathbb{Z}\right\}$, where $\Delta\tau=\tau_{\mathrm{p}}/M$ and $\Delta\nu=\nu_{\mathrm{p}}/N$ denote the delay and Doppler resolutions, respectively. The corresponding frame duration and bandwidth are given by $T=N\tau_{\mathrm{p}}$ and $B=M\nu_{\mathrm{p}}$, respectively.
The $M\times N$ DD grid within the fundamental region corresponds to $k=0,\ldots,M-1$ and $l=0,\ldots,N-1$. Let $x[k,l]\in\mathbb{A}$, $k=0,\ldots,M-1$ and $l=0,\ldots,N-1$, denote the $MN$ information symbols placed on this grid, where $\mathbb{A}$ is the modulation alphabet. The symbols are extended over the entire DD plane according to the quasi-periodicity condition as
\begin{equation}
x_{\rm dd}[k+nM,l+mN]
=
x[k,l]e^{j2\pi nl/N},
\qquad n,m\in\mathbb{Z}.
\end{equation}
The corresponding continuous DD domain signal is 
constructed as
\vspace{0mm}
\begin{equation}
x_{\rm dd}(\tau,\nu)
=
\sum_{k,l\in\mathbb{Z}}
x_{\rm dd}[k,l]
\delta(\tau-k\Delta\tau)\delta(\nu-l\Delta\nu),
\end{equation}
which satisfies 
$x_{\rm dd}(\tau+n\tau_{\mathrm{p}},\nu+m\nu_{\mathrm{p}})
=
e^{j2\pi n\nu\tau_{\mathrm{p}}}x_{\rm dd}(\tau,\nu)$,
where $n,m\in\mathbb{Z}$,
and is therefore periodic along the Doppler dimension and quasi-periodic along the delay dimension. 

To obtain a waveform with prescribed time and bandwidth containment, 
a DD domain pulse shaping is applied using a separable transmit filter
$w_{\rm tx}(\tau,\nu)
=
w_\tau(\tau)w_\nu(\nu)$,
where $w_\tau(\tau)$ and $w_\nu(\nu)$ denote the pulse shaping functions along the delay and Doppler dimensions, respectively. The pulse-shaped DD domain transmit signal is given by
\begin{equation}
x_{\rm dd}^{w_{\rm tx}}(\tau,\nu)
=
w_{\rm tx}(\tau,\nu)*_{\sigma}x_{\rm dd}(\tau,\nu),
\end{equation}
where $*_{\sigma}$ denotes twisted convolution\footnote{Twisted convolution of two DD domain functions $a(\tau,\nu)$ and $b(\tau,\nu)$ is defined as
$a(\tau,\nu)*_{\sigma}b(\tau,\nu)
\triangleq
\int_{-\infty}^{\infty}\int_{-\infty}^{\infty}
a(\tau',\nu')b(\tau-\tau',\nu-\nu')
e^{j2\pi\nu'(\tau-\tau')}
\,d\tau'\,d\nu'$.
The twisted convolution operation preserves quasi-periodicity.}.
The corresponding time domain transmit signal is obtained through inverse Zak transform\footnote{Inverse Zak transform of a DD domain signal $a(\tau,\nu)$ is defined as $Z_{t}^{-1}\{a(\tau,\nu)\} \triangleq \sqrt{\tau_{\mathrm p}} \int_0^{\nu_{\mathrm p}} a(t,\nu) d\nu$.} as
$s_{\rm td}(t)
=
\mathcal{Z}^{-1}_{t}
\left\{
x_{\rm dd}^{w_{\rm tx}}(\tau,\nu)
\right\}$.
The transmitted signal propagates through a doubly-dispersive wireless channel consisting of $P$ propagation paths, whose DD domain representation is modeled as
\begin{equation}
\label{h_phy}
h(\tau,\nu)
=
\sum_{i=1}^{P}
h_i\,
\delta(\tau-\tau_i)
\delta(\nu-\nu_i),
\end{equation}
where $h_i$, $\tau_i$, and $\nu_i$ denote the complex gain, delay, and Doppler of the $i$th path, respectively. The received time domain signal at the receiver is given by
$y_{\rm td}(t)=r_{\rm td}(t)+n_{\rm td}(t)$, where
$r_{\rm td}(t)=
\iint
h(\tau,\nu)
s_{\rm td}(t-\tau)
e^{j2\pi\nu(t-\tau)}
\,d\tau d\nu$
denotes the channel output, and $n_{\rm td}(t)$ denotes additive white Gaussian noise (AWGN) with variance $N_0$, characterized by
$\mathbb{E}\!\left[n_{\rm td}(t)n_{\rm td}^{*}(t+t')\right]
=N_0\delta(t')$. The received time domain signal $y_{\rm td}(t)$ is transformed into DD domain by applying Zak transform\footnote{Zak transform of a continuous time domain signal $a(t)$ is defined as
$Z_t\{a(t)\} \triangleq \sqrt{\tau_{\mathrm{p}}}
\sum_{k\in\mathbb{Z}}a(\tau+k\tau_{\mathrm p})
e^{-j2\pi\nu k\tau_{\mathrm p}}$.} as
\begin{equation}
y_{\rm dd}(\tau,\nu)= \mathcal{Z}_{t}\{y_{\rm td}(t)\} = r_{\rm dd}(\tau,\nu)+n_{\rm dd}(\tau,\nu),
\label{eqydd}
\end{equation}
where $r_{\rm dd}(\tau,\nu)=\mathcal{Z}_{t}\{r_{\rm td}(t)\}$ and
$n_{\rm dd}(\tau,\nu)=\mathcal{Z}_{t}\{n_{\rm td}(t)\}$. The DD domain signal in (\ref{eqydd}) is further processed using a receive pulse shaping filter $w_{\rm rx}(\tau,\nu)$, which is matched to the transmit filter according to
$w_{\rm rx}(\tau,\nu)=w_{\rm tx}^{*}(-\tau,-\nu)e^{j2\pi\tau\nu}$.
It operates on the received DD signal through twisted convolution, yielding the DD input-output (I/O) relation as
\begin{align}
y_{\rm dd}^{w_{\rm rx}}(\tau,\nu)
={}&
w_{\rm rx}(\tau,\nu)*_{\sigma}y_{\rm dd}(\tau,\nu)
\nonumber\\
={}&
\underbrace{
w_{\rm rx}(\tau,\nu)*_{\sigma}
h(\tau,\nu)*_{\sigma}
w_{\rm tx}(\tau,\nu)
}_{\triangleq\,h_{\rm eff}(\tau,\nu)}
*_{\sigma}x_{\rm dd}(\tau,\nu)
\nonumber\\
&+
\underbrace{
w_{\rm rx}(\tau,\nu)*_{\sigma}
n_{\rm dd}(\tau,\nu)
}_{\triangleq\,n_{\rm dd}^{w_{\rm rx}}(\tau,\nu)},
\label{cont_io}
\end{align}
where $h_{\rm eff}(\tau,\nu)$ denotes the effective DD domain channel formed by the twisted convolution cascade of the transmit filter, the physical channel, and the receive filter, while
$n_{\rm dd}^{w_{\rm rx}}(\tau,\nu)$ denotes the corresponding receive-filtered DD domain noise. The filtered DD domain signal is sampled on the DD information lattice to obtain the discrete DD domain signal
$y^{w_{\text{rx}}}_{\rm dd}[k,l]
=
y_{\rm dd}^{w_{\rm rx}}
\left(k\Delta\tau,l\Delta\nu\right)$,
where $k,l\in\mathbb{Z}$.
The resulting discrete DD domain I/O relation can be expressed as
\begin{equation}
y^{w_{\text{rx}}}_{\rm dd}[k,l]
=
h_{\rm eff}[k,l]*_{\sigma_{\rm d}}x_{\rm dd}[k,l]
+
n_{\rm dd}[k,l],
\end{equation}
where $*_{\sigma_{\rm d}}$ denotes the discrete twisted convolution
operation\footnote{The discrete twisted convolution of two DD domain sequences $a[k,l]$ and $b[k,l]$ is defined as
$a[k,l]*_{\sigma_{\rm d}}b[k,l]
\triangleq
\sum_{k',l'\in\mathbb{Z}}
a[k-k',l-l']b[k',l']
e^{j2\pi k'(l-l')/(MN)}$.},
$h_{\rm eff}[k,l] 
=
h_{\rm eff}
\left(\tau = k\Delta\tau, \nu = l\Delta\nu\right)$,
and
$n_{\rm dd}[k,l] 
=
n_{\rm dd}^{w_{\rm rx}}
\left(\tau = k\Delta\tau, \nu = l\Delta\nu\right)$. As a consequence of quasi-periodicity, it is sufficient to consider the received samples within the fundamental region $\mathcal{D}_0$. Stacking these $MN$ samples into vectors, the end-to-end DD domain I/O relation can be written as
\begin{equation}
\mathbf{y}
=
\mathbf{H}_{\rm eff}\mathbf{x}
+
\mathbf{n},
\label{sys_mod}
\end{equation}
where $\mathbf{x},\mathbf{y},\mathbf{n}\in\mathbb{C}^{MN\times1}$, with $x_{kN+l+1}=x_{\rm dd}[k,l]$ denoting the $(kN+l+1)$th entry of $\mathbf{x}$, and similarly for $\mathbf{y}$ and $\mathbf{n}$. The effective channel matrix $\mathbf{H}_{\rm eff}\in\mathbb{C}^{MN\times MN}$ has entries
\begin{multline}
\mathbf{H}_{\rm eff}[k'N+l'+1,kN+l+1]
=
\\
\sum_{m,n\in\mathbb{Z}}
h_{\rm eff}[\Delta k,\Delta l]
e^\frac{j2\pi nl}{N}
e^{j2\pi\frac{\Delta l(k+nM)}{MN}},
\label{eqn_channel_matrix}
\end{multline}
where $\Delta k=k'-k-nM$ and $\Delta l=l'-l-mN$, with $k,k'=0,\ldots,M-1$ and $l,l'=0,\ldots,N-1$.

\vspace{-2mm}
\subsection{Model-Free Channel Estimation}
A key advantage of Zak-OTFS is that it enables simple, model-free channel estimation when operated in the crystalline regime\footnote{The system is said to be operating in the crystalline regime when the delay and Doppler spreads of the effective DD domain channel are smaller than the delay and Doppler periods, respectively, i.e., $\tau_{\rm eff}<\tau_{\mathrm{p}}$ and $\nu_{\rm eff}<\nu_{\mathrm{p}}$.}. Under this condition, the effective channel response is well confined within a dedicated pilot region, allowing the effective I/O relation to be directly read off from the received samples in the pilot region, denoted by $y_{\mathrm{P}}[k,l]$, as~\cite{zak_otfs2}
\begin{equation}
\widehat{h}_{\mathrm{eff}}[k,l]=
\begin{cases}
y_{\mathrm{P}}\!\left[k+\frac{M}{2},\,l+\frac{N}{2}\right] e^{-j\pi \frac{l}{N}}, & \hspace{-2mm} -k_L \le k < k_R, \\
& -\frac{N}{2}\le l < \frac{N}{2}, \\
0, & \hspace{-2mm} \text{otherwise,}
\end{cases}
\label{hatheff}
\end{equation} 
where $k_L$ and $k_R$ denote the left and right delay-domain boundaries of the pilot region in an embedded frame, chosen to capture the effective channel spread. The estimates of $h_{\mathrm{eff}}[k,l]$s in (\ref{hatheff}) are used in (\ref{eqn_channel_matrix}) to obtain estimates of the entries of the $\mathbf{H}_{\rm eff}$ matrix. The resulting $\widehat{\mathbf{H}}_{\rm eff}$ matrix is used for subsequent equalization/detection. This model-free estimation is quite simple and it eliminates the need to estimate the individual physical channel parameters (i.e., path gains, delays, and Dopplers $\{h_i, \tau_i, \nu_i\}_{i=1}^P$) using complex model-dependent channel estimation algorithms.

\subsection{Benchmark DD Filters and Their 64-QAM Performance}
\label{sub:filters}
Several DD domain pulse shaping filters, including sinc, Gaussian, root raised cosine (RRC), GS, Hermite, and isotropic orthogonal transform algorithm (IOTA) filters \cite{zak_otfs2},\cite{zak_otfs5}-\cite{zak_otfs8}, have been considered in the Zak-OTFS literature. In this paper, we consider sinc, Gaussian, and GS filters for benchmarking purposes. The sinc and Gaussian pulses represent different ends of the orthogonality-localization tradeoff. The sinc pulse satisfies the Nyquist criterion with nulls at non-zero DD grid-points, but its sidelobes decay slowly resulting in poor DD localization. In contrast, the Gaussian pulse provides good DD localization with no sidelobes, but lacks nulls at non-zero DD grid-points causing ISI. These characteristics influence the channel estimation and data detection performance. The GS pulse combines the favorable properties of both Gaussian and sinc pulses, though its effective pulse spread remains significantly broader than that of the Gaussian pulse.

\emph{Sinc pulse:}
The sinc pulse is given by
\begin{equation}
w_{\mathrm{tx}}(\tau,\nu)
=
\sqrt{B}\,\mathrm{sinc}(B\tau)
\sqrt{T}\,\mathrm{sinc}(T\nu).
\label{eq:sinc}
\end{equation}

\emph{Gaussian pulse:}
The Gaussian pulse is given by
\begin{equation}
w_{\mathrm{tx}}(\tau,\nu)
=
\Big(\frac{2\alpha_{\tau}B^{2}}{\pi}\Big)^{\frac{1}{4}}
e^{-\alpha_{\tau}B^{2}\tau^{2}}
\Big(\frac{2\alpha_{\nu}T^{2}}{\pi}\Big)^{\frac{1}{4}}
e^{-\alpha_{\nu}T^{2}\nu^{2}},
\label{eq:gauss}
\end{equation}
where $\alpha_{\tau}$ and $\alpha_{\nu}$ are the roll-off parameters for controlling localization along the delay and Doppler dimensions, respectively. Since the Gaussian pulse has infinite support, its effective time duration and bandwidth are defined using a $99\%$ energy-containment criterion, which corresponds to  
$\alpha_{\tau}=\alpha_{\nu}=1.584$. 

\emph{GS pulse:}
The GS pulse is given by
\begin{align}
w_{\mathrm{tx}}(\tau,\nu) = 
& \hspace{1mm}
\Omega_{\tau}\Omega_{\nu}\sqrt{BT}\,
\mathrm{sinc}(B\tau)\mathrm{sinc}(T\nu) \nonumber \\
& \hspace{1mm}
e^{-\alpha_{\tau}B^{2}\tau^{2}}
e^{-\alpha_{\nu}T^{2}\nu^{2}},
\label{eq:GS}
\end{align}
where $\alpha_{\tau}$ and $\alpha_{\nu}$ are the roll-off parameters and $\Omega_{\tau}$ and $\Omega_{\nu}$ are the corresponding energy-normalization factors. For no time or bandwidth expansion, $\alpha_{\tau}=\alpha_{\nu}=0.044$ and $\Omega_{\tau}=\Omega_{\nu}=1.0278$~\cite{zak_otfs6}. 

\begin{figure}
\centering
\includegraphics[width=0.95\linewidth,height=0.75\linewidth]{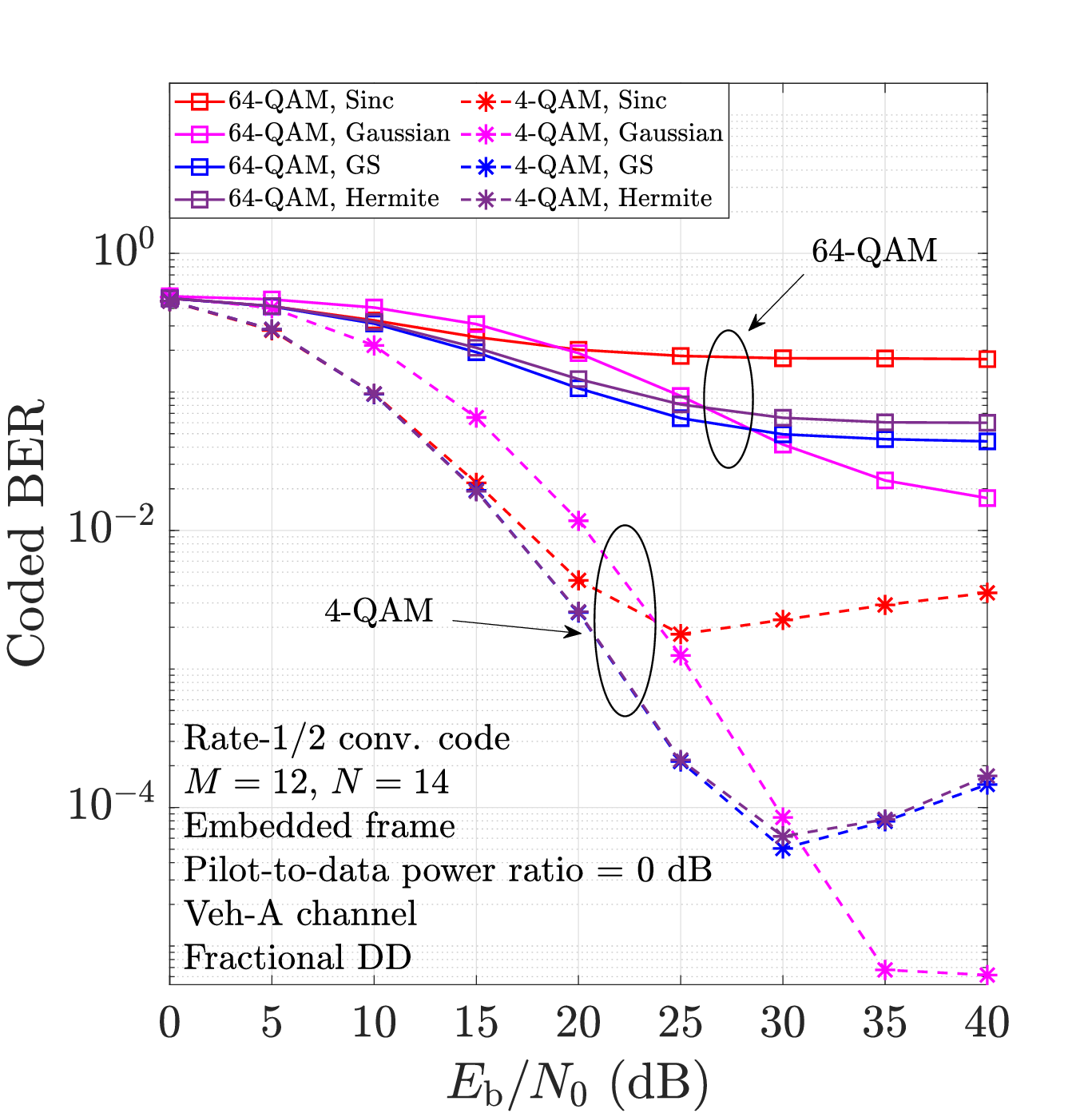}
\caption{Coded BER vs $E_\text{b}/N_0$ performance of benchmark sinc, Gaussian, GS, and Hermite filters for 4-QAM and 64-QAM.}
\vspace{-4mm}
\label{fig:coded_benchmark}
\end{figure}

{\em BER performance for 64-QAM:}
We note that most works in the Zak-OTFS literature have mainly considered lower-order modulation such as BPSK, 4-QAM, and 8-QAM. There has not been much focus on higher-order modulation, which is of interest. We observe that while the considered filters in the literature have shown good performance with lower-order modulation, their performance with higher-order modulation is poor. We illustrate this in Fig.~\ref{fig:coded_benchmark} which shows the coded BER performance of Zak-OTFS with sinc, Gaussian, GS, and Hermite\footnote{The Hermite filter is constructed as in \cite{zak_otfs7}. As established therein, we used $N_c=9$ which yields the best performance among the considered choices of $N_c$, and its performance is on par with that of the GS filter, which is also observed in Fig. \ref{fig:coded_benchmark}. Accordingly, we omit  Hermite filter in subsequent figures to avoid overcrowding of plots, as its performance is represented by GS filter performance. We also do not consider RRC filter, as it involves bandwidth/time expansion.} filters for 4-QAM and 64-QAM with rate-1/2 convolutional code,  embedded pilot frame with 0 dB pilot-data power ratio, model-free channel estimation, and MMSE detection. As expected, 4-QAM performance is better than 64-QAM performance. The increased BER for large $E_\mathrm{b}/N_0$ in 4-QAM can be attributed to channel-response truncation arising from the limited pilot region in the embedded frame, along with residual pilot-data interference. Importantly, for 64-QAM, the filters exhibit significantly poor performance with floors at high error rates ($>10^{-2}$ coded BER). 
Motivated by this, in the following section, we develop a filter under the MM filter design framework which achieves a significantly better BER performance for higher-order QAM. 

\begin{figure*}[t]
\centering
\begin{subfigure}[b]{0.32\textwidth}
    \includegraphics[width=\textwidth,height=0.75\textwidth]{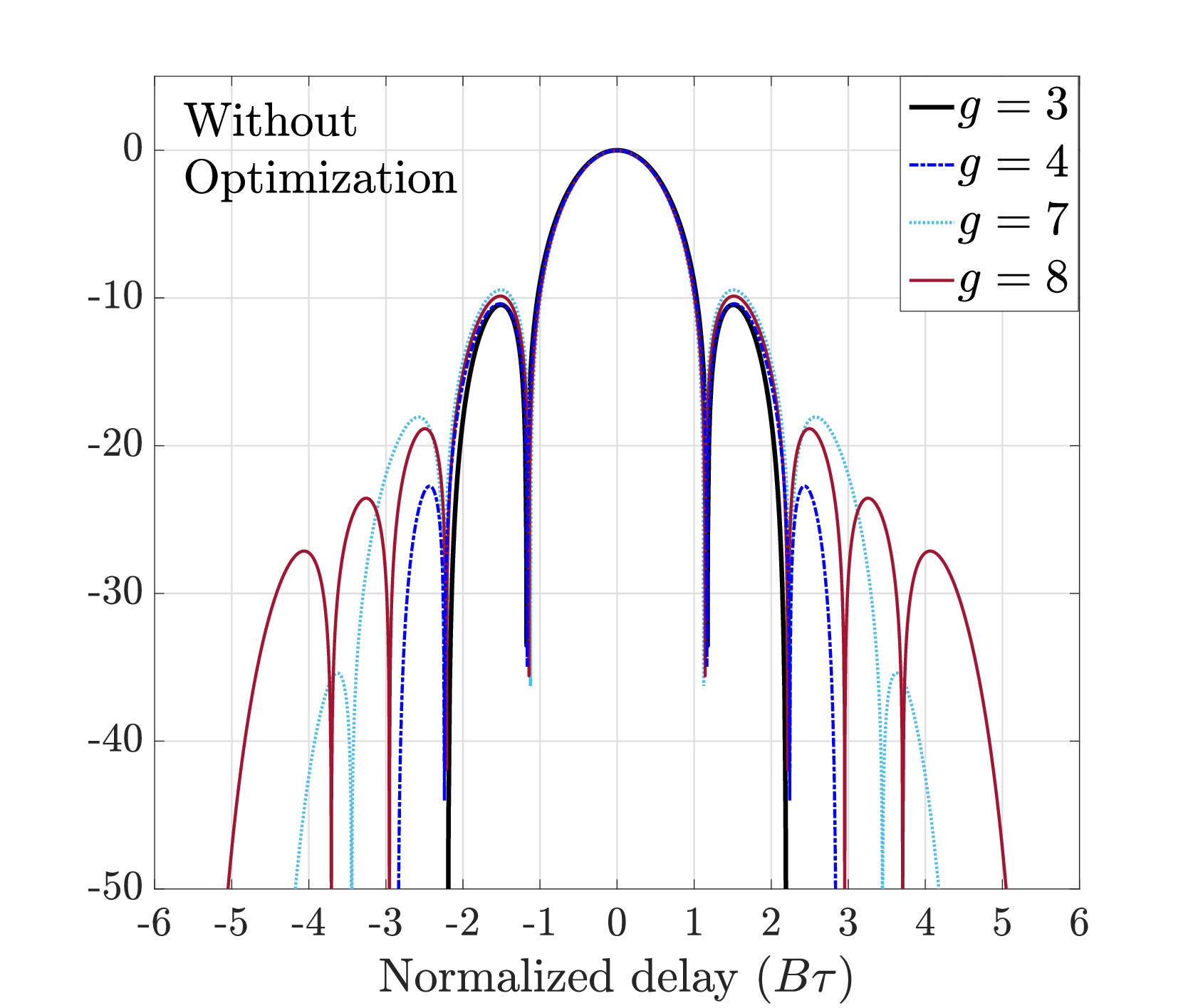}
    \subcaption{}
\end{subfigure}
\begin{subfigure}[b]{0.32\textwidth}
    \includegraphics[width=\textwidth,height=0.75\textwidth]{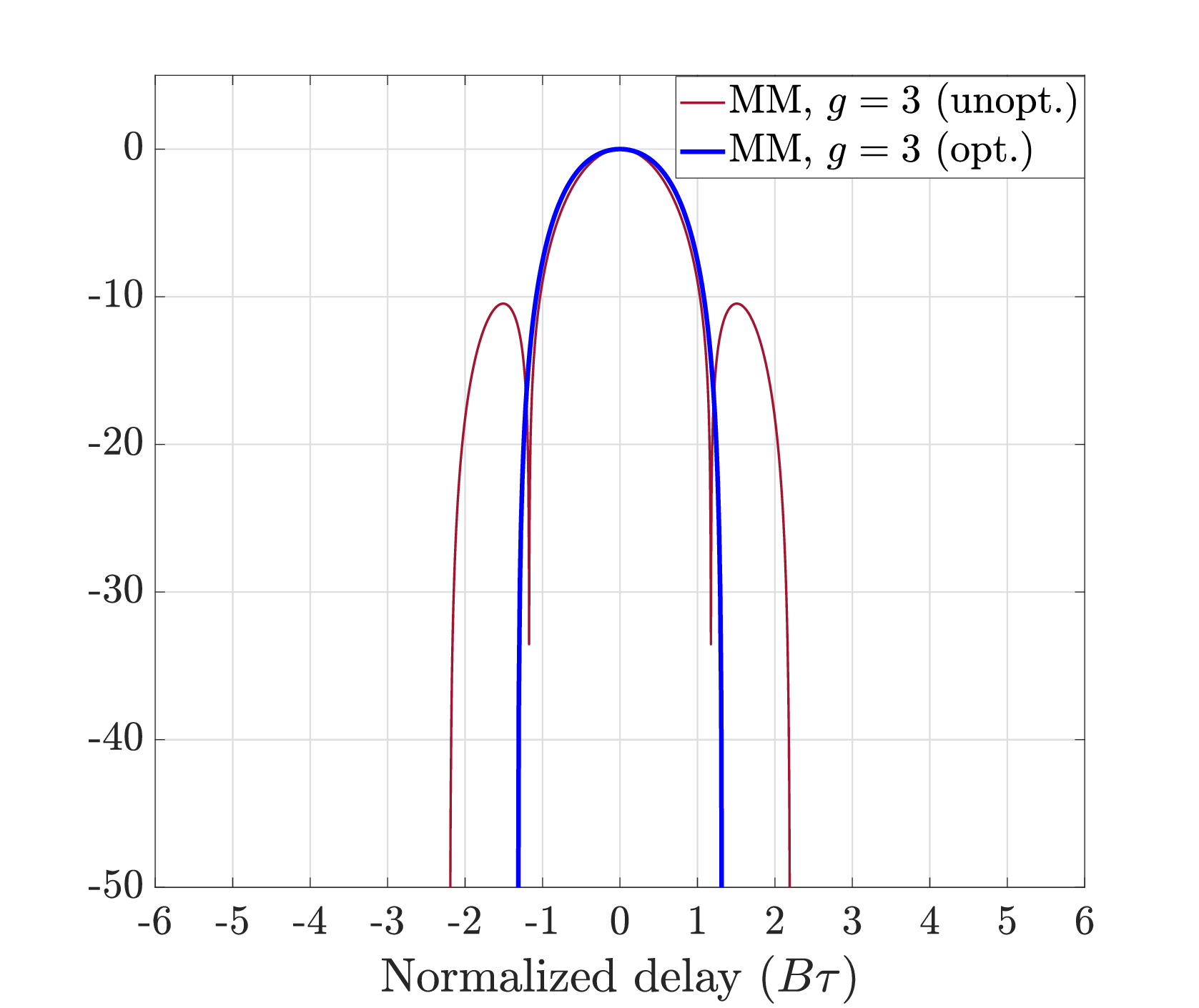}
    \subcaption{}
\end{subfigure}
\begin{subfigure}[b]{0.32\textwidth}
    \includegraphics[width=\textwidth,height=0.75\textwidth]{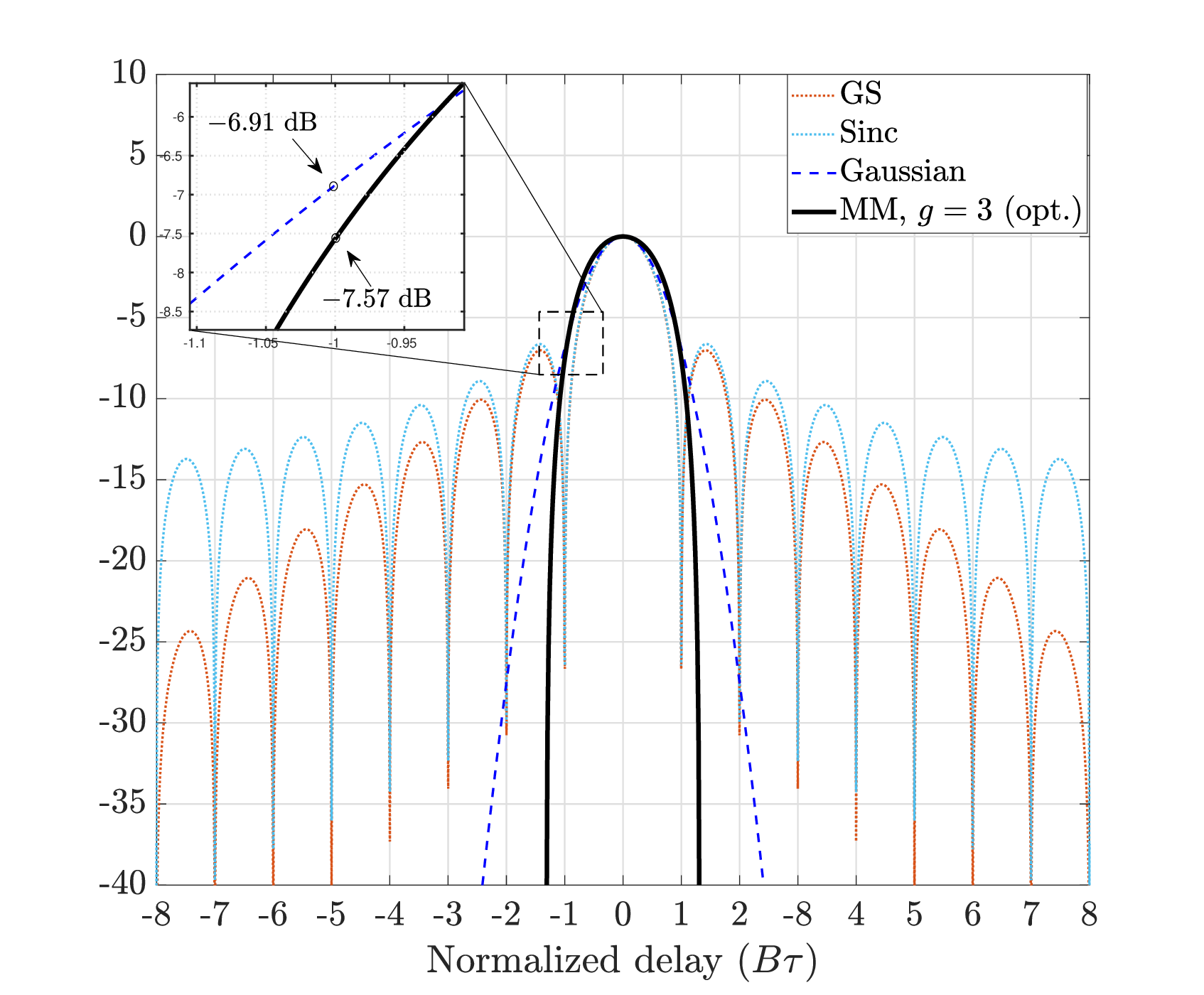}
    \subcaption{}
\end{subfigure}
\vspace{-1mm}
\caption{Comparison of delay-domain MM pulse magnitudes, $10\log_{10}|w_1(\tau)|$ (dB): (a) unoptimized MM pulses for $g=3$, $4$, $7$, $8$; (b) MM pulse for $g=3$ before and after coefficient optimization; (c) sinc, Gaussian, GS, and optimized $g=3$ MM pulses.}
\label{fig:comp_pulse_mag}
\vspace{-2mm}
\end{figure*}

\section{Proposed MM Filter for Zak-OTFS}
\label{sec:proposed}
In this section, we present the construction of the proposed two-dimensional MM pulse for Zak-OTFS, its characteristics, and the optimization of the pulse coefficients.

\subsection{Construction of the Proposed Two-Dimensional MM Pulse}
\label{sub:design}
The proposed pulse is constructed by adapting the one-dimensional prototype filter proposed in~\cite{MM_1} to the two-dimensional DD domain. In the original filter construction, the integer overlap factor $g$ determines the length of the prototype filter and, consequently, the extent over which adjacent symbol waveforms overlap in time. The MM design provides a distinct set of prototype coefficients for each value of $g$; in particular, the original work considers $3\leq g\leq 8$ and tabulates the corresponding coefficient sets together with their stopband characteristics~\cite{MM_1}. Thus, different values of $g$, together with their corresponding coefficient sets, generate a family of MM pulses with different time-domain durations and frequency-domain localization characteristics.

For a given $g$, the corresponding MM coefficient set is used to construct one-dimensional pulses along the delay and Doppler dimensions. These pulses are respectively written as
\begin{align}
p_{\tau}(\tau)
&=
c_0+
2\sum_{\ell=1}^{g-1}c_{\ell}
\cos\!\left[
\frac{2\pi\ell}{\zeta_{\tau}}
\left(B\tau+\frac{\zeta_{\tau}}{2}\right)
\right],
\label{p_tau}\\
p_{\nu}(\nu)
&=
c_0+
2\sum_{\ell=1}^{g-1}c_{\ell}
\cos\!\left[
\frac{2\pi\ell}{\zeta_{\nu}}
\left(T\nu+\frac{\zeta_{\nu}}{2}\right)
\right],
\label{p_nu}
\end{align}
where $\{c_0,c_1,\ldots,c_{g-1}\}$ are the MM coefficients associated with the selected value of $g$, and $\zeta_{\tau}$ and $\zeta_{\nu}$ are the corresponding roll-off factors that control the pulse extents in the delay and Doppler dimensions, respectively. These parameters are selected such that $99\%$ of the pulse energy is contained within the prescribed time duration $T$ and bandwidth $B$, respectively. The corresponding one-dimensional delay and Doppler domain pulses are given by
\begin{align}
w_{\tau}(\tau)
&=
\kappa_{\tau}\sqrt{B}\,
p_{\tau}(\tau)
\operatorname{rect}\!\left(\frac{B\tau}{\zeta_{\tau}}\right),
\label{w_tau}\\
w_{\nu}(\nu)
&=
\kappa_{\nu}\sqrt{T}\,
p_{\nu}(\nu)
\operatorname{rect}\!\left(\frac{T\nu}{\zeta_{\nu}}\right),
\label{w_nu}
\end{align}
where
$\kappa_{\tau}$ and $\kappa_{\nu}$ are normalization factors chosen such that $w_{\tau}(\tau)$ and $w_{\nu}(\nu)$, respectively, have unit energy. The resulting two-dimensional DD transmit pulse is then formed as the separable product $w_{\tau}(\tau)w_{\nu}(\nu)$, given by
\begin{equation}
w_{\mathrm{tx}}(\tau,\nu)
=
\kappa_{\tau}
\kappa_{\nu}\sqrt{BT}\,
p_{\tau}(\tau)p_{\nu}(\nu)
\operatorname{rect}\!\left(\frac{B\tau}{\zeta_{\tau}}\right)
\operatorname{rect}\!\left(\frac{T\nu}{\zeta_{\nu}}\right).
\label{w_tx}
\end{equation}

\subsection{Characteristics of the Two-Dimensional MM Pulse Family}
\label{sub:char}
\begin{table}
\centering
\begin{tabular}{|c||c|c|c|c|c|}
\hline
Metric & $g=3$ & $g=4$ & $g=7$ & $g=8$ \\ \hline\hline
$W_{\mathrm{@-40\,dB}}$ & 4.35 & 5.61 & 7.84 & 9.66  \\ \hline
$P_{\mathrm{@\,max\text{-}SL}}$ (dB) & -10.46 & -10.39 & -9.45 & -9.87 \\ \hline
$P_{@\,1/B}$ (dB) & -8.96 & -9.21 & -10.30 & -9.79 \\ \hline
$P_{@\,2/B}$ (dB) & -18.24 & -15.78 & -14.29 & -15.09 \\ \hline
\end{tabular}
\caption{Delay-domain characteristics of unoptimized MM pulses for different $g$.}
\vspace{-4mm}
\label{tab:un_opt}
\end{table}

The effect of the overlap factor $g$ on the pulse shape is illustrated in Fig.~\ref{fig:comp_pulse_mag}(a), which shows the delay-domain pulse magnitude for $g=3$, $4$, $7$, and $8$ using their corresponding unoptimized MM coefficient sets $\{c_0,c_1,\ldots,c_{g-1}\}$ reported in Table~I of \cite{MM_1}. Since the proposed DD pulse is separable and its delay and Doppler components have identical functional forms, the Doppler-domain pulses exhibit a similar trend and are therefore omitted for brevity. For clarity, only the four cases of $g=3,4,7,8$ are shown to avoid overlapping curves (a similar trend is observed for the other values of $g$). To ensure a fair comparison across different values of $g$, the roll-off factors $\zeta_{\tau}$ and $\zeta_{\nu}$ are selected according to the common $99\%$ energy-containment criterion, without time or bandwidth expansion. It can be observed that as $g$ increases, the pulse becomes increasingly spread out.

To quantitatively compare the pulse characteristics, we define four quantities. The first quantity is the pulsewidth at $-40$~dB  pulse magnitude, denoted by $W_{\mathrm{@-40\,dB}}$, defined as the distance between the outermost $-40$ dB crossings, beyond which the pulse magnitude remains below $-40$~dB. The width is normalized by $1/B$. The second quantity is the maximum sidelobe magnitude, denoted by $P_{\mathrm{@\,max\text{-}SL}}$, defined as the maximum pulse magnitude in the sidelobe region outside the central main lobe. The third and fourth quantities are $P_{\mathrm{@\,1/B}}$ and $P_{\mathrm{@\,2/B}}$, which denote the pulse magnitudes at the first and second non-zero DD grid-points, i.e., at $\tau=1/B$ and $\tau=2/B$, respectively. These four quantities for different values of $g$ are summarized in Table~\ref{tab:un_opt}. As can be observed from the table, the pulse with $g=3$ has the smallest $-40$ dB width among the considered values of $g$, while the maximum sidelobe and DD grid-point values remain comparable across the different choices of $g$. Hence, increasing $g$ leads to a broader pulse without providing a corresponding improvement in the suppression of DD grid-point values. This makes the pulse with $g=3$ particularly attractive for the DD domain setting, where compactness is desirable while maintaining small values at the DD grid-points. This raises the question of whether the MM pulse can be further tailored to the DD domain requirements through coefficient optimization while retaining the underlying MM structure.

\vspace{-1mm}
\subsection{Optimization of MM Pulse Coefficients}
\label{sub:opt}
The original MM coefficients are designed to exploit increased temporal overlap with larger $g$ to achieve improved frequency-domain selectivity, including high stopband attenuation and fast sidelobe fall-off, while satisfying the near-perfect reconstruction requirement of the underlying filter-bank system~\cite{MM_1}. In contrast, the proposed pulse is constructed directly in the DD domain, where localization is constrained by the $99\%$ energy-containment criterion within the prescribed $T$ and $B$. Thus, the frequency-domain tradeoff underlying the original MM design is not the primary design objective here. Instead, the MM coefficients are further optimized to adapt the pulse to the DD domain requirements while retaining the underlying MM structure as follows.

For a given value of $g$, we denote the corresponding MM coefficient vector by\footnote{The delay and Doppler domain components have the same normalized MM functional form, differing only in their respective dimensional scaling through $B\tau$ and $T\nu$. Hence, the coefficients are optimized in the delay domain and the resulting coefficient vector is used for both dimensions, with $\zeta_{\tau}$ and $\zeta_{\nu}$ selected independently for the respective energy-containment requirements.}
\begin{equation}
\mathbf{c}_{\mathrm{MM}}^{(g)}
=
[c_0,c_1,\ldots,c_{g-1}]^{T}.
\end{equation}
These original coefficients serve as the starting point for the proposed DD domain optimization. The first coefficient, $c_{\mathrm{MM},0}^{(g)}=1$, is fixed at unity, which normalizes the filter response at zero frequency without loss of generality \cite{MM_1}. Specifically, we consider a modified coefficient vector
\begin{equation}
\mathbf{c}^{(g)}
=
\mathbf{c}_{\mathrm{MM}}^{(g)}
+
\Delta\mathbf{c}^{(g)},
\end{equation}
where
$\Delta\mathbf{c}^{(g)}
=
[0,\Delta c_1,\ldots,\Delta c_{g-1}]^{T}$
denotes the coefficient perturbation, ensuring that $c_0^{(g)}=1$ remains unchanged.

For each candidate coefficient vector, the roll-off parameters are determined numerically to satisfy the prescribed energy-containment requirements. Specifically, $\zeta_{\tau}$ and $\zeta_{\nu}$ are selected so that $99\%$ of the pulse energy is contained within the prescribed bandwidth and time-duration intervals, $[-B/2,B/2]$ and $[-T/2,T/2]$, respectively. Thus, each candidate pulse is evaluated under the same time and bandwidth constraints, allowing the coefficient optimization to focus on improving DD domain localization without expanding the prescribed support.

To quantify localization, we define {\em sidelobe energy loss}, which calculates the aggregate energy of the delay-domain pulse over a specified sidelobe region:
\begin{equation}
J_{\mathrm{SL}}^{(g)}\!\big(\mathbf{c}^{(g)}\big)
=
\sum_{\tau\in\mathcal{R}_{\mathrm{SL}}^{(g)}}
\left|w_{\tau}(\tau)\right|^2,
\label{eq:J_SL}
\end{equation}
where $\mathcal{R}_{\mathrm{SL}}^{(g)}$ denotes the positive sidelobe region bounded by $\tau_{\mathrm{L}}^{(g)}$ and $\tau_{\mathrm{U}}^{(g)}$, with sampling points separated by $\Delta_{\mathrm{SL}}$. Here, $\tau_{\mathrm{L}}^{(g)}$ separates the main lobe region from the sidelobe region, while $\tau_{\mathrm{U}}^{(g)}$ is chosen to include the relevant sidelobe energy.

\begin{figure}[t]
\centering
\includegraphics[width=0.95\linewidth,height=0.75\linewidth]{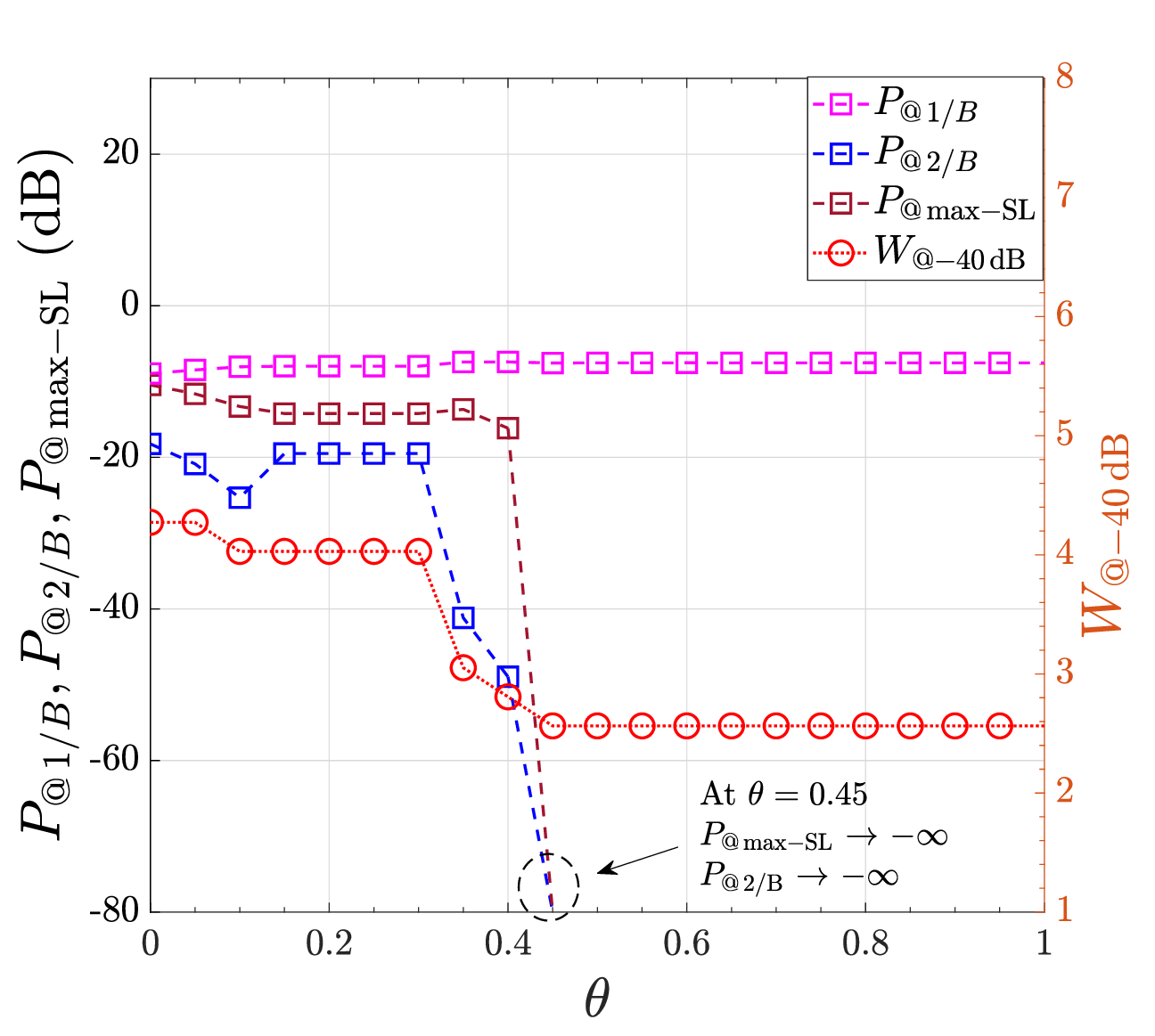}
\caption{Performance metric comparison for different $\theta$ values.}
\vspace{-4mm}
\label{fig:delta}
\end{figure}

The coefficient optimization minimizes $J_{\mathrm{SL}}^{(g)}\!\left(\mathbf{c}^{(g)}\right)$ while
constraining the optimized coefficients to remain within a prescribed
distance of their original MM values. The resulting problem is
formulated as
\begin{equation}
\begin{aligned}
\underset{\mathbf{c}^{(g)}}{\operatorname{minimize}}
\quad & J_{\mathrm{SL}}^{(g)}\!\big(\mathbf{c}^{(g)}\big)\\
\text{subject to}\quad
& c_0^{(g)}=1,\\
& \big|c_i^{(g)}-c_{\mathrm{MM},i}^{(g)}\big|
\leq \theta, \quad i=1,\ldots,g-1.
\end{aligned}
\label{eq:opt}
\end{equation}
Here, $\theta$ controls the admissible coefficient perturbations and therefore the extent to which the original MM prototype can be modified. Its value is selected empirically by sweeping candidate
bounds and comparing the resulting pulse characteristics, $P_{\mathrm{@\,max\text{-}SL}}, P_{\mathrm{@\,1/B}}, P_{\mathrm{@\,2/B}}$, and $W_{\mathrm{@-40\,dB}}$, as shown in Fig.~\ref{fig:delta}. This comparison determines a suitable bound that improves localization while retaining desirable characteristics of the original prototype.

In the present work, we specialize this general framework to the $g=3$ MM pulse. We select $g=3$ as the starting point because it exhibits the most compact DD domain structure among the considered unoptimized MM pulses, as observed in Table~\ref{tab:un_opt}. This compactness is desirable for model-free I/O relation estimation, as it confines the effective channel response more tightly around the pilot. At the same time, the responses at non-zero DD grid-points  influence interference between neighboring data symbols. The design therefore seeks to reduce sidelobe energy while retaining sufficiently low responses at these grid-points, rather than imposing exact nulls as explicit optimization constraints\footnote{Although exact grid-point nulls are desirable for orthogonality, enforcing them does not minimize the aggregate sidelobe energy. The present optimization prioritizes pulse localization, allowing a tradeoff between exact grid-point nulling and concentration of energy around the main lobe.}. 

For the $g=3$ pulse, the first zeros occur close to $\pm 1/B$, while the second zeros occur near $\pm 2/B$, which approximately defines the overall extent of the pulse. Based on these observations, we choose the sidelobe region as
$[\tau_{\mathrm{L}}^{(3)},\tau_{\mathrm{U}}^{(3)}]=[1/B,2/B]$,
with a sampling interval
$\Delta_{SL}=\frac{5\times10^{-3}}{B}$ for evaluating $J_{\mathrm{SL}}^{(3)}$. Based on the empirical comparison in Fig.~\ref{fig:delta}, we select $\theta=0.45$ as it yields a substantial reduction in $W_{\mathrm{@-40\,dB}}$ and suppresses $P_{\mathrm{@\,max\text{-}SL}}$ and $P_{@\,2/B}$ to $-\infty$ dB, while maintaining a relatively low $P_{@\,1/B}$. This choice therefore provides a favorable balance between pulse localization and the responses at non-zero grid-points. With these choices, the constrained optimization over $\Delta\mathbf{c}^{(3)}$ is solved numerically using sequential quadratic programming (SQP). Starting from the original MM coefficient vector
\begin{equation}
\mathbf{c}_{\mathrm{MM}}^{(3)}=[1,\,-0.91143,\,0.41143]^T,
\end{equation}
the constrained optimization yields the coefficient vector
\begin{equation}
\mathbf{c}_{\mathrm{opt}}^{(3)}
=
[1,\,-0.46143,\,-0.03856]^T.
\label{eq:opt_coeff}
\end{equation}
The corresponding optimized roll-off factors are $\zeta_\tau=\zeta_\nu=2.6320$, with normalization factors $\kappa_\tau=\kappa_\nu=0.5156$. Table~\ref{tab:opt_params} summarizes the original and optimized MM pulse parameters for comparison.

\begin{figure}
\centering \includegraphics[width=1.0\linewidth]{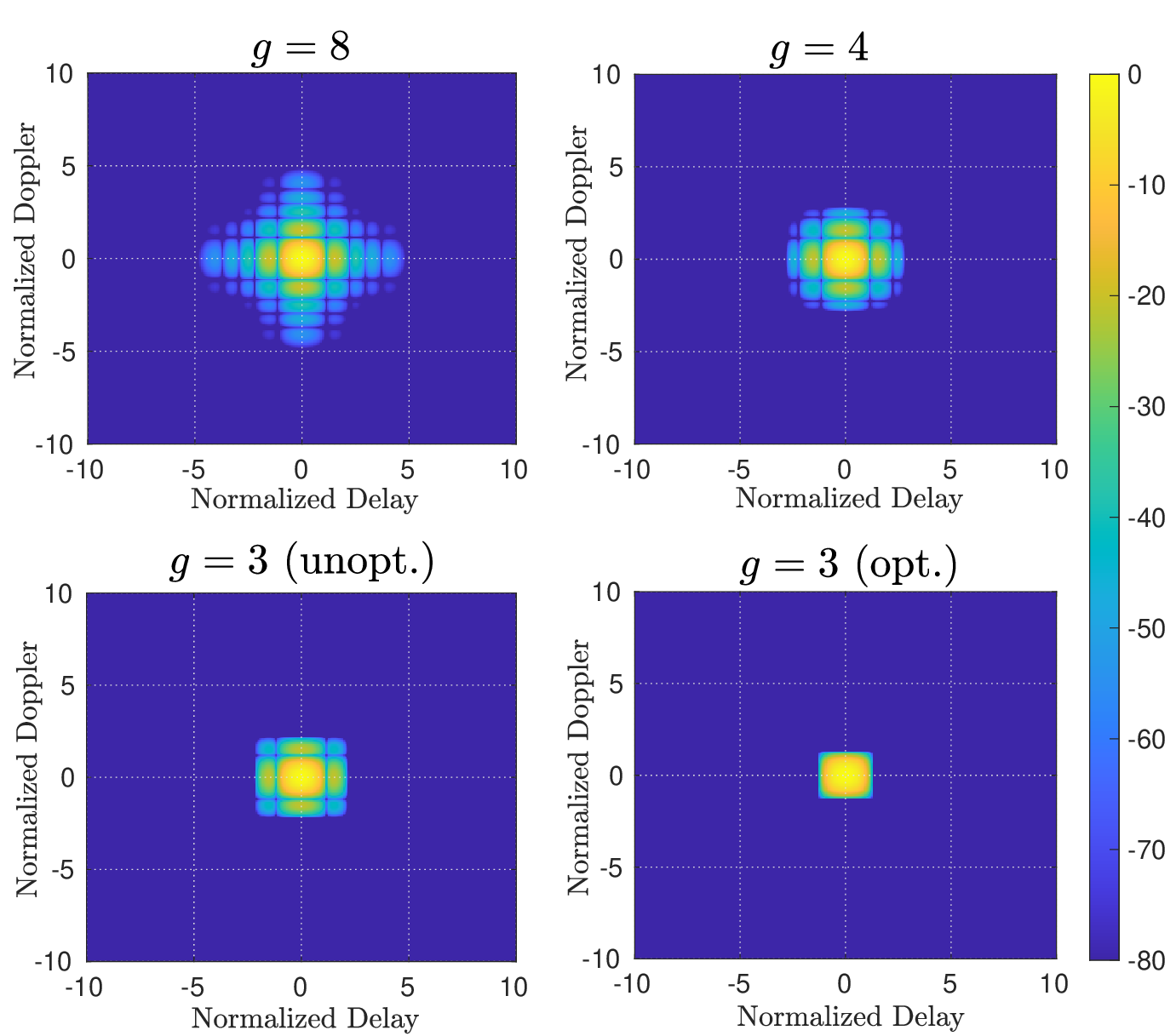}
\caption{Comparison of the heatmaps of the MM pulse, $w_{\mathrm{tx}}(\tau,\nu)$, for unoptimized pulses with $g=3$, $4$, and $8$, and the optimized pulse with $g=3$.}
\label{fig:heatmap}
\end{figure}

\begin{table}[t]
\centering
\begin{tabular}{|c||c|c|}
\hline
Parameter & MM (unopt.) & MM (opt.) \\ \hline\hline
$\mathbf{c}^{(3)}$
& $[1,-0.91143,0.41143]^T$
& $[1,-0.46143,-0.03856]^T$ \\ \hline
$\zeta_\tau=\zeta_\nu$ & 4.3913 & 2.6320 \\ \hline
$\kappa_\tau=\kappa_\nu$ & 0.2755 & 0.5156 \\ \hline
\end{tabular}
\caption{Unoptimized and optimized parameters for the $g=3$ MM pulse.}
\label{tab:opt_params}
\vspace{-5mm}
\end{table}

\vspace{-3mm}
\subsection{Comparison of Optimized MM and Benchmark Pulses}
\label{sub:opt_char}
The effect of coefficient optimization on the $g=3$ MM pulse is illustrated in Fig.~\ref{fig:comp_pulse_mag}(b), which compares the optimized (opt.) pulse with its unoptimized (unopt.) counterpart. The optimized MM pulse exhibits no sidelobes, resulting in significantly improved localization compared with its unoptimized counterpart. To further illustrate its energy distribution, Fig.~\ref{fig:heatmap} compares the DD domain magnitude heatmaps of the optimized $g=3$ pulse with those of the unoptimized MM pulses for $g\in\{3,4,8\}$. The heatmaps highlight the enhanced compactness of the optimized pulse, with its energy more tightly concentrated around the origin and minimal energy spreading into the surrounding DD domain.

We next compare the optimized MM pulse with the sinc, Gaussian, and GS pulses in terms of their magnitude responses, as shown in Fig.~\ref{fig:comp_pulse_mag}(c). The sinc pulse exhibits exact nulls at the DD grid-points but has a large pulse spread, 
whereas the Gaussian pulse is more compact but exhibits higher responses at the first and second DD grid-points, potentially leading to increased ISI. The GS pulse combines Gaussian shaping with the grid-null property of the sinc pulse, but its pulse spread remains larger than those of the Gaussian and optimized MM pulses. In comparison, the optimized MM pulse achieves the most compact response, with no sidelobes and a lower response at the first DD grid-point than the Gaussian pulse, offering a favorable balance between localization and detection performance.

These observations are quantified in Table~\ref{tab:pulse_char}, which compares the pulsewidth and sidelobe characteristics of the considered pulses. The optimized MM pulse achieves the smallest $W_{\mathrm{@-40\,dB}}$, reducing the pulsewidth from $4.35$ to $2.61$ relative to the unoptimized MM pulse. Furthermore, both $P_{\mathrm{@\,max\text{-}SL}}$ and $P_{\mathrm{@\,2/B}}$ are reported as $-\infty$ dB, while $P_{\mathrm{@\,1/B}}$ is lower than that of the Gaussian pulse, indicating improved suppression of the response at the first DD grid-point. These results confirm the improved compactness and sidelobe suppression achieved through coefficient optimization without explicitly enforcing DD grid-point nulls.

\section{Closed-Form Expressions for Effective Channel and Noise Covariance}
\label{sec:closed_form}
In this section, we derive closed-form expressions for the effective DD domain channel and the covariance of the filtered DD domain noise for the proposed MM filter. These expressions reduce the computational time required to evaluate the effective channel and noise statistics, which in turn reduce the simulation run times significantly.

\begin{table}
\setlength{\tabcolsep}{3pt}
\centering
\begin{tabular}{|c||c|c|c|c|c|}
\hline
Metric & Sinc & Gaussian & GS & MM (unopt.) & MM (opt.) \\ \hline\hline
$W_{\mathrm{@-40\,dB}}$ & 6365 & 4.82 & 26.98 & 4.35 & 2.61 \\ \hline
$P_{\mathrm{@\,max\text{-}SL}}$ (dB) & $-6.63$ & $-$ & $-6.90$ & $-10.46$ & $-$ \\ \hline
$P_{@\,1/B}$ (dB) & $-\infty$ & $-6.91$ & $-\infty$ & $-8.96$ & $-7.57$ \\ \hline
$P_{@\,2/B}$ (dB) & $-\infty$ & $-27.52$ & $-\infty$ & $-18.24$ & $-\infty$ \\ \hline
\end{tabular}
\caption{Comparison of normalized delay-domain pulse characteristics.}
\vspace{-4mm}
\label{tab:pulse_char}
\end{table}

\begin{theorem}
\label{thm:closed_form_heff}
The DD domain effective channel $h_{\mathrm{eff}}(\tau,\nu)$ for the proposed MM filter is given by 
\begin{eqnarray}
\hspace{-6mm}
h_{\mathrm{eff}}(\tau,\nu) & = & 
\kappa_{\tau}^2 \,\kappa_{\nu}^2 \,B \,T\,
\sum_{i=1}^{P}\, h_i
\,\,\Lambda_{\tau,i}(\tau)
\,\,\Lambda_{\nu,i}(\nu)
\nonumber \\
& & \hspace{-5mm} \times
\sum_{m,n,r,s=-(g-1)}^{g-1}
\hspace{-2mm} a^*_m\,a_n\,a^*_r\, a_s\,\,
e^{j\Phi_{mnrs}^{(i)}(\tau,\nu)}
\nonumber \\
& & \hspace{-5mm} \times
\hspace{1mm} \operatorname{sinc}(\beta_{mn}^{(i)}\Lambda_{\tau,i}(\tau))
\operatorname{sinc}(\gamma_{rs}(\tau)\Lambda_{\nu,i}(\nu)),
\label{eq:heff_cf}
\end{eqnarray}
where
\begin{align}
\Lambda_{\tau,i}(\tau)
&\triangleq
\Big[
\frac{\zeta_{\tau}}{B}
-
\left|\tau-\tau_i\right|
\Big]^{+},
\label{eq:Lambda_tau}
\\
\Lambda_{\nu,i}(\nu)
&\triangleq
\Big[
\frac{\zeta_{\nu}}{T}
-
\left|\nu-\nu_i\right|
\Big]^{+},
\label{eq:Lambda_nu}
\end{align}
\vspace{0mm}
with $[x]^+\triangleq\max(x,0)$, and
\begin{align}
\beta_{mn}^{(i)}
&\triangleq
\frac{(m-n)B}{\zeta_{\tau}}
-\nu_i,
\label{eq:beta_mn}
\\
\gamma_{rs}(\tau)
&\triangleq
\tau+
\frac{(r-s)T}{\zeta_{\nu}}.
\label{eq:gamma_rs}
\end{align}
The phase term is
\begin{align}
\Phi_{mnrs}^{(i)}(\tau,\nu)
&\triangleq
\pi
\left(
\nu_i+
\frac{(m+n)B}{\zeta_{\tau}}
\right)
(\tau-\tau_i)
\nonumber\\
&+
\pi
\left(
\tau+
\frac{(r+s)T}{\zeta_{\nu}}
\right)
(\nu-\nu_i).
\label{eq:Phi_mnrs}
\end{align}
The finite-series coefficients are given by
\begin{equation}
a_z
=
\begin{cases}
c_0, & z=0,\\
(-1)^{|z|}c_{|z|}, & 1\le |z|\le g-1,
\end{cases}
\qquad
z\in\{m,n,r,s\}.
\label{eq:a_q}
\end{equation}
\end{theorem}

\begin{IEEEproof}
See Appendix \ref{app:heff}.
\end{IEEEproof}

\begin{theorem}
\label{thm:noise_cov}
For the proposed MM filter, the covariance between the DD domain noise samples at $(k_1,l_1)$ and $(k_2,l_2)$ is given by
\begin{align}
&\mathbb{E}\!\left[
n_{\mathrm{dd}}[k_1,l_1]n_{\mathrm{dd}}^{*}[k_2,l_2]
\right] \nonumber\\
&\quad =
N_0\tau_{\mathrm{p}}
\sum_{q_1,q_2\in\mathbb Z}
e^{-j2\pi(q_1l_1-q_2l_2)/N}
\mathcal{G}_{\nu}(x_1)\mathcal{G}_{\nu}^{*}(x_2)
C_{\tau}(\Delta_x),
\label{eq:noise_cov}
\end{align}
where
\begin{equation}
x_i=q_i\tau_{\mathrm{p}}+k_i\Delta\tau,
\qquad i\in\{1,2\},
\label{eq:noise_x}
\end{equation}
and
\begin{equation}
\Delta_x\triangleq x_2-x_1.
\label{eq:noise_delta_x}
\end{equation}
The Doppler-domain factor $\mathcal{G}_{\nu}(x)$ is given by
\begin{equation}
\mathcal{G}_{\nu}(x)
=
\kappa_{\nu}\frac{\zeta_{\nu}}{\sqrt{T}}
\sum_{z=-(g-1)}^{g-1}
a_z\,
\operatorname{sinc}\!\left(
\frac{\zeta_{\nu}x}{T}+z
\right),
\label{eq:noise_Gnu}
\end{equation}
while the delay-domain correlation factor is
\begin{align}
C_{\tau}(\Delta_x)
&=
\kappa_{\tau}^{2}B\,
\Gamma_{\tau}(\Delta_x)
\sum_{m,n=-(g-1)}^{g-1}
a_m^{*}a_n
\nonumber\\
&\quad{}\times
e^{-j\frac{\pi B}{\zeta_{\tau}}
(m+n)\Delta_x}
\operatorname{sinc}\!\left(
\frac{B}{\zeta_{\tau}}(m-n)
\Gamma_{\tau}(\Delta_x)
\right),
\label{eq:noise_Ctau}
\end{align}
with the overlap length
\begin{equation}
\Gamma_{\tau}(\Delta_x)
\triangleq
\left[
\frac{\zeta_{\tau}}{B}
-|\Delta_x|
\right]^{+},
\label{eq:noise_Gamma}
\end{equation}
where the finite-series coefficients $a_z$ are defined in \eqref{eq:a_q}.
\end{theorem}

\begin{IEEEproof}
See Appendix \ref{app:noise_cov}.
\end{IEEEproof}

Although \eqref{eq:noise_cov} is formally expressed over $q_1, q_2 \in \mathbb{Z}$, it effectively reduces to a finite sum in implementation. Specifically, because the delay-domain pulse has finite support, the overlap length $\Gamma_{\tau}(\Delta_x)$ (and hence $C_{\tau}(\Delta_x)$) vanishes whenever $|\Delta_x| \geq \zeta_{\tau}/B$, which restricts non-zero contributions to closely spaced values of $q_1$ and $q_2$. Furthermore, since $\mathcal{G}_{\nu}(x_i)$ decays rapidly as $|x_i|$ grows with $|q_i|$, only a few terms around $q_1, q_2 = 0$ contribute non-negligibly to the sum.

{\em Note on simulation run time:} 
For the system parameters considered in Fig. \ref{fig:nmse}, evaluating the effective DD domain channel over $100$ channel realizations using the closed-form expression requires approximately $7.71$~s, compared with $253.78$~s using numerical integration. For the computation of the noise covariance matrix, the closed-form expression requires approximately $14.24$~s, compared with $290.50$~s using numerical integration. The simulations are performed using MATLAB R2024b on a system equipped with an Intel\textsuperscript{\textregistered} Core\textsuperscript{\texttrademark} Ultra 9 285K processor with 24 cores, an NVIDIA GeForce RTX 4060 GPU with 8~GB memory, and 192~GB RAM.

\section{Results and Discussions}
\label{sec:results}
In this section, we present simulation results to illustrate the performance of the proposed MM filter in comparison with those of sinc, Gaussian, GS filters for different system settings.

\subsection{Simulation Parameters}
We evaluate the performance of the proposed MM filter in Zak-OTFS over  Vehicular A (Veh-A) channel \cite{veh}, 
consisting of $P=6$ paths with fractional delays and Dopplers, 
and power-delay profile given in Table~\ref{tab_veh}. The maximum delay spread is $\tau_{\max}=2.51~\mu$s and the maximum Doppler shift is $\nu_{\max}=815$~Hz. The Doppler shift of the $i$th path is generated as $\nu_i=\nu_{\max}\cos(\phi_i)$, where $\phi_i$ is
independently and uniformly distributed over $[0,2\pi)$. A Doppler period of $\nu_{\mathrm{p}}=15$ kHz and the corresponding delay period of $\tau_{\mathrm{p}} = 1/ \nu_{\mathrm{p}} = 66.67~\mu$s are considered. The delay period is sliced into $M=12$ bins with delay resolution $1/B$ and the Doppler period is sliced into $N=14$ bins with Doppler resolution $1/T$, which correspond to a bandwidth of $B=M\nu_{\mathrm{p}}=180$ kHz and a frame duration of $T=N\tau_{\mathrm{p}}=0.93$ ms. The proposed MM pulse with $g=3$ and the optimized coefficient vector in \eqref{eq:opt_coeff} is used throughout the simulations. All BER results are obtained using the minimum mean-square error (MMSE) detector. Unless otherwise specified, the above configuration is used for the simulations.

\subsection{Embedded Pilot Configuration}
\begin{figure}[t]
\centering
\includegraphics[width=0.95\linewidth,height=0.7\linewidth]{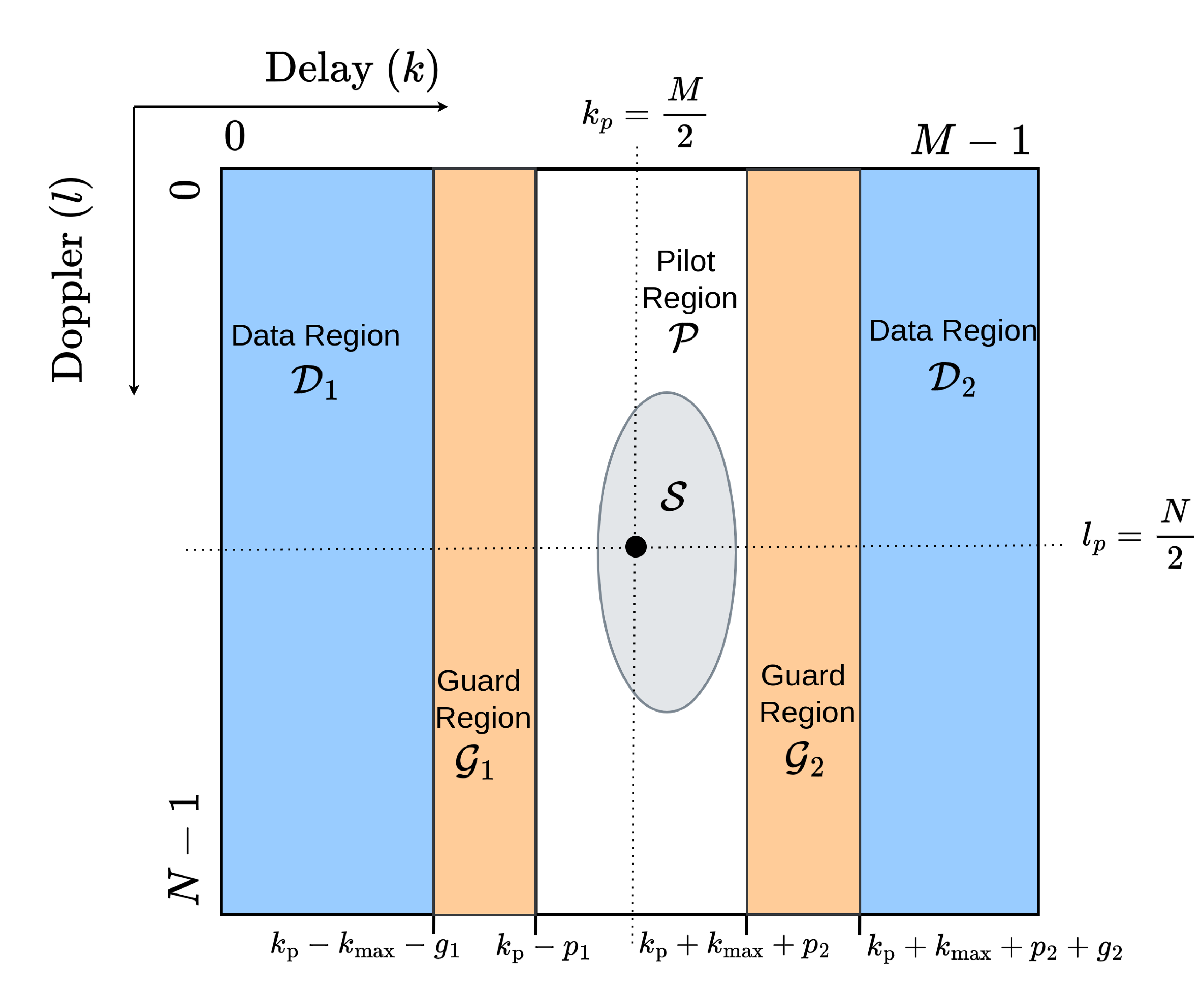}
\caption{Embedded pilot frame in Zak-OTFS.}   
\label{fig:embedd}
\end{figure}

Model-free I/O relation estimation is performed using the embedded
pilot configuration shown in Fig.~\ref{fig:embedd}. The pilot is placed at $(k_p,l_p)=\left(\frac{M}{2},\frac{N}{2}\right)$. Since the effective DD domain channel is spread due to the physical channel delay and Doppler spreads as well as the pulse shaping, the effective channel response is obtained by reading the received samples over a region around the pilot. Specifically, the pilot region $\mathcal{P}$ spans from $k_L = k_p-p_1$ to $k_R = k_p+k_{\max}+p_2$, where $k_{\max}=\lceil B\tau_{\max}\rceil$ and $\tau_{\max}\triangleq\max(\tau_i)$, with $p_1$ and $p_2$ denoting
non-negative integers that provide additional bins to capture the
pulse-induced spread. The regions $\mathcal{G}_1$ and $\mathcal{G}_2$
on either side of $\mathcal{P}$ are defined by
$k_p-k_{\max}-g_1$ to $k_p-p_1-1$ and
$k_p+k_{\max}+p_2+1$ to $k_p+k_{\max}+p_2+g_2$, respectively, where
$g_1$ and $g_2$ are non-negative integers that provide separation
between the pilot and data regions and thereby limit pilot-to-data
interference. The remaining DD grid-points, 
$\mathcal{D}=\mathcal{D}_1\cup\mathcal{D}_2$, are used for data 
symbols. The symbols within the embedded DD frame are allocated as
\vspace{0mm}
\begin{equation}
x[k,l]=
\begin{cases}
\sqrt{E_\mathrm{p}}, & (k,l)=(k_\mathrm{p},l_\mathrm{p}), \\
\sqrt{\frac{E_\mathrm{d}}{|\mathcal{D}|}}x_{\mathrm{d}}[k,l], & (k,l)\in \mathcal{D}, \\
0, & \mathrm{otherwise},
\end{cases}
\end{equation}
where $x_\mathrm{d}[k,l]$ denotes the data symbol with $\mathbb{E}[|x_\mathrm{d}[k,l]|^{2}]=1$, $E_\mathrm{p}$ is the energy allocated to the pilot symbol, and $E_\mathrm{d}$ is the total energy allocated to all data symbols across the frame. The average energy per frame is given by $E_\mathrm{p}+E_\mathrm{d}$. For the $M=12$, $N=14$ DD frame, we set $p_1=0$, $p_2=2$, $g_1=0$, and $g_2=1$. With $B=180$~kHz and $\tau_{\max}=2.51~\mu$s, the maximum delay index is $k_{\max}=\lceil B\tau_{\max}\rceil=1$. The frame utilization is defined as $\eta\triangleq|\mathcal{D}|/(MN)$, where $|\mathcal{D}|$ denotes the number of DD-grid locations allocated for data symbols. 
For this configuration, $|\mathcal{D}|=84$, yielding $\eta=50\%$.

\begin{table}
\centering
\begin{tabular}{|c|c|c|c|c|c|c|}
\hline
Path index ($i$)         & 1 & 2    & 3    & 4    & 5    & 6    \\ \hline
Delay $\tau_{i}$ ($\mu s$)      & 0 & 0.31 & 0.71 & 1.09 & 1.73 & 2.51 \\ \hline
Relative power (dB) & 0 & -1   & -9   & -10  & -15  & -20  \\ \hline
\end{tabular}
\caption{Power delay profile of Vehicular-A channel model.}
\label{tab_veh}
\vspace{-2mm}
\end{table}
The bit-energy-to-noise-density ratio is given by
\begin{equation}
\frac{E_\mathrm{b}}{N_0}
\triangleq
\frac{E_\mathrm{d}}
{|\mathcal{D}|\,\log_2(Q)\,R_c\,N_0},
\label{eq:snr}
\end{equation}
where $Q=|\mathbb{A}|$ is the modulation order, and $R_c=k_c/n_c$ is the code rate, with $k_c$ and $n_c$ denoting the numbers of information and coded bits, respectively. For the uncoded system, $R_c=1$. The pilot-to-data power ratio (PDR) is defined as
\begin{equation}
\mathrm{PDR}\triangleq\frac{E_\mathrm{p}}{E_\mathrm{d}},
\end{equation}
where a PDR of $0$~dB corresponds to equal power allocation between the pilot symbol and all data symbols ($E_\mathrm{p} = E_\mathrm{d}$). For all simulations, the PDR is set to $0$~dB unless otherwise specified.

\subsection{Performance Comparison With Benchmark Filters}

\begin{figure}
\centering
\includegraphics[width=0.95\linewidth,height=0.75\linewidth]{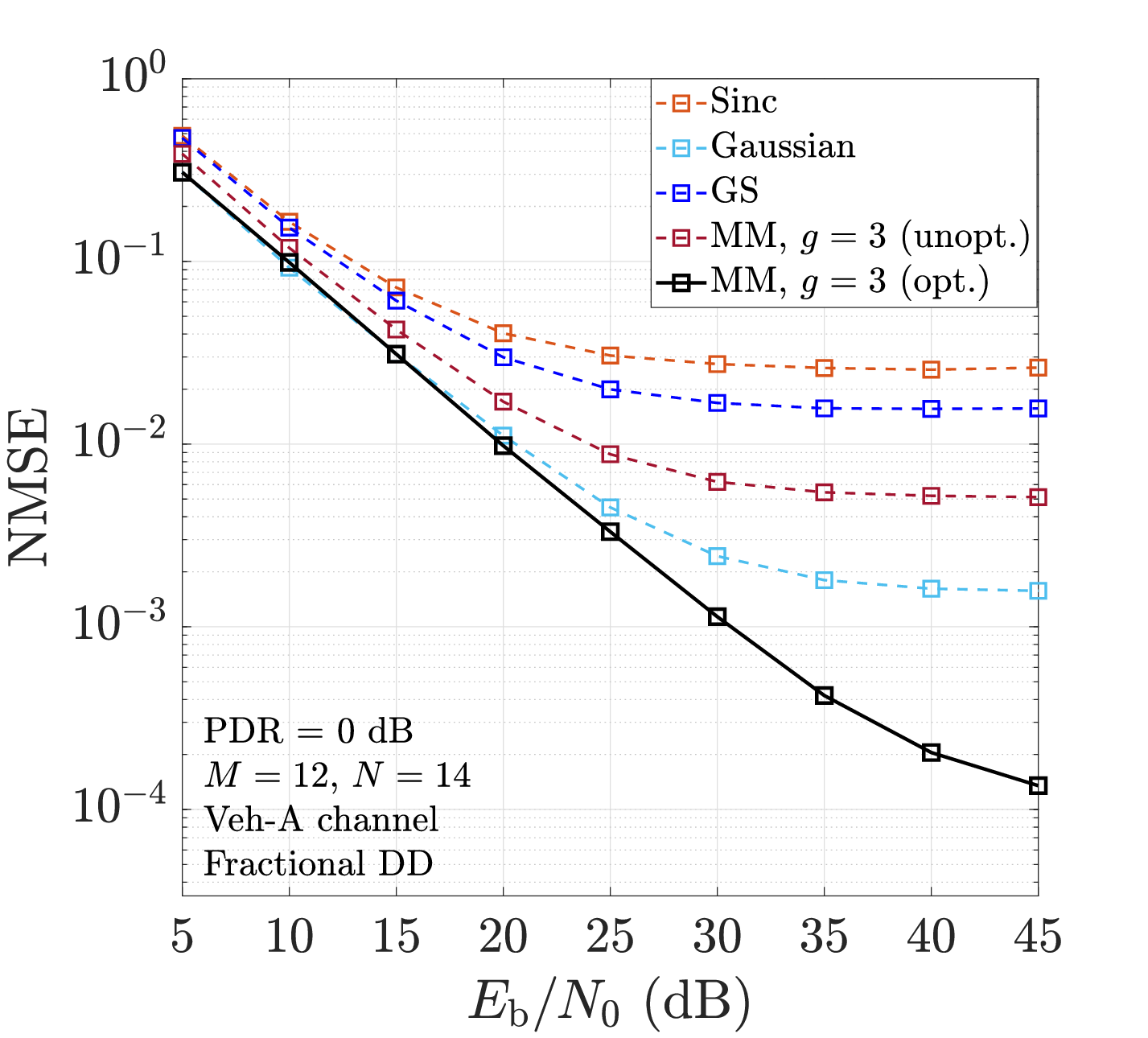}
\caption{NMSE vs $E_\text{b}/N_0$ performance of different filters.} 
\label{fig:nmse}
\vspace{-4mm}
\end{figure}

{\em NMSE Performance:} Fig.~\ref{fig:nmse} shows the I/O relation estimation performance of different filters in terms of NMSE as a function of $E_\text{b}/N_0$. The NMSE is evaluated between the estimated and the true effective DD domain channel matrices as
$(\|\widehat{\mathbf{H}}_{\mathrm{eff}}
-\mathbf{H}_{\mathrm{eff}}\|_{\mathrm{F}}^{2})
/{\|\mathbf{H}_{\mathrm{eff}}\|_{\mathrm{F}}^{2}}$. As the $E_\text{b}/N_0$ increases, the NMSE improves for all the considered pulses. The unoptimized MM filter with $g=3$ provides better NMSE compared to the sinc and GS filters, but remains inferior compared to the Gaussian filter, which provides strong DD domain localization. In contrast, the optimized MM filter substantially improves the NMSE performance, outperforming the Gaussian filter. 
This improvement is mainly due to the optimized coefficients, which transform the original MM pulse into a more compact, sidelobe-free response, resulting in better DD domain localization and more accurate I/O relation estimation.

\begin{figure*}[t]
\centering
\begin{subfigure}[b]{0.32\textwidth}
    \includegraphics[width=\textwidth,height=0.75\textwidth]{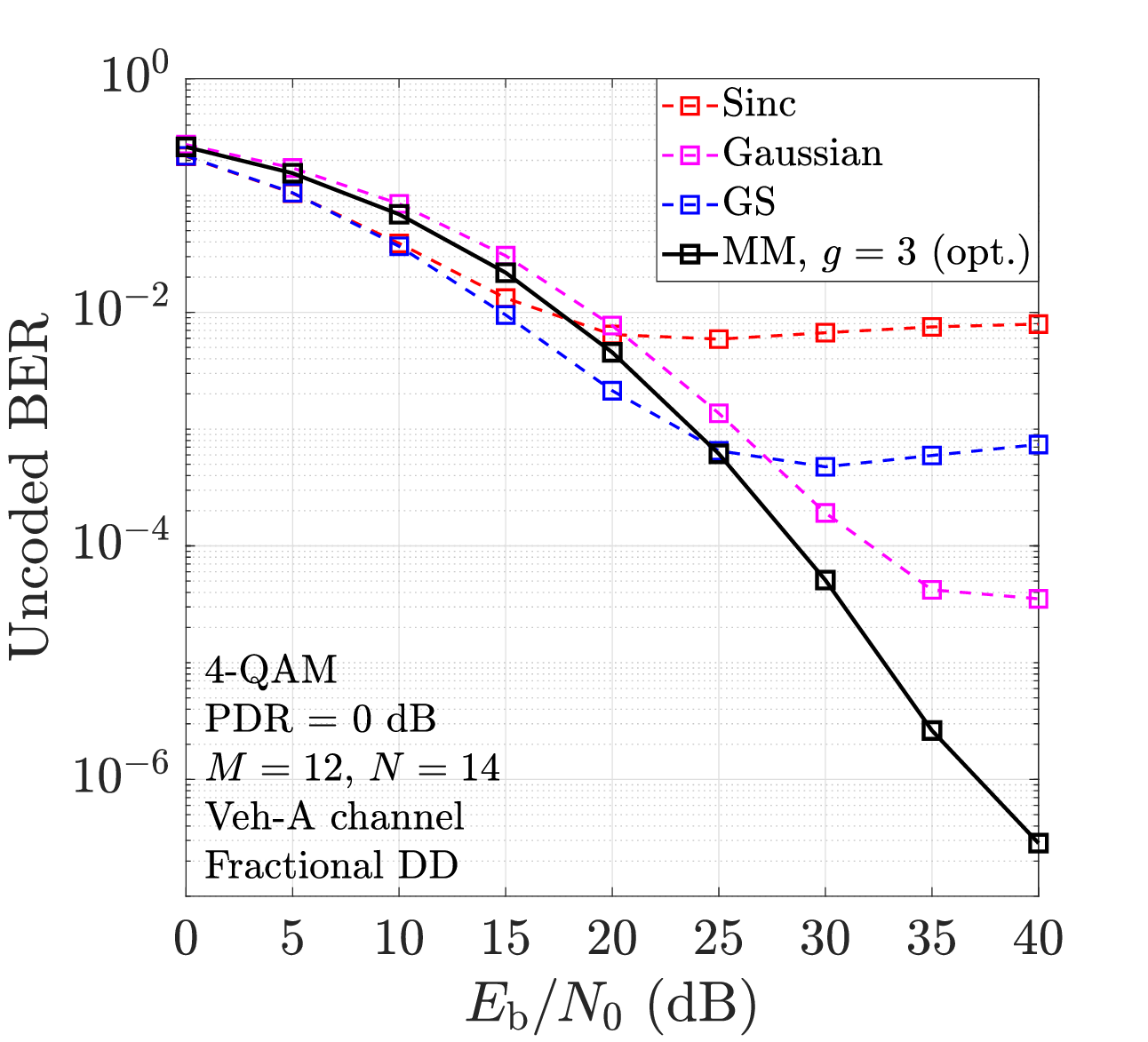}
    \subcaption{}
\end{subfigure}
\begin{subfigure}[b]{0.32\textwidth}
    \includegraphics[width=\textwidth,height=0.75\textwidth]{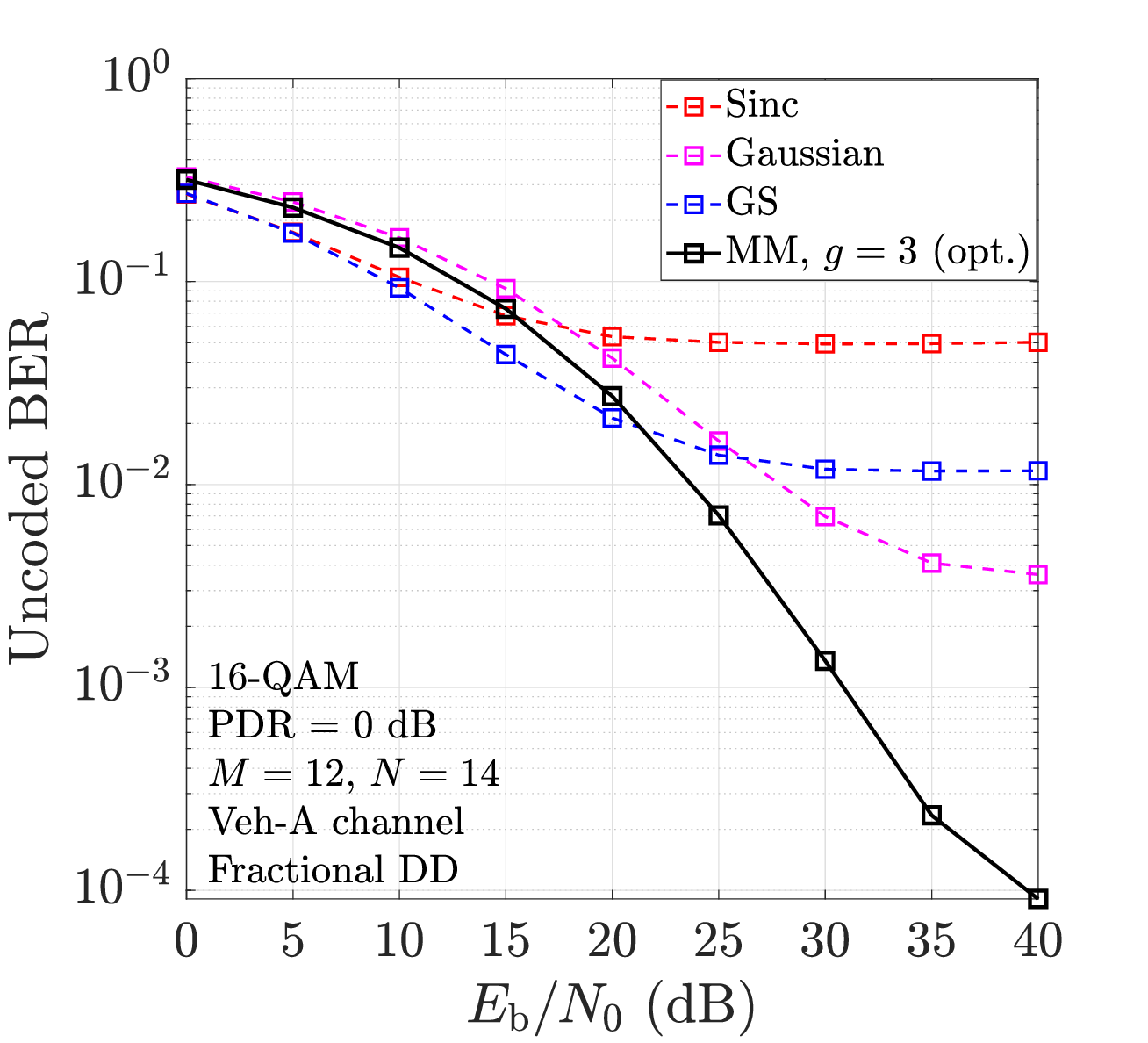}
    \subcaption{}
\end{subfigure}
\begin{subfigure}[b]{0.32\textwidth}
    \includegraphics[width=\textwidth,height=0.75\textwidth]{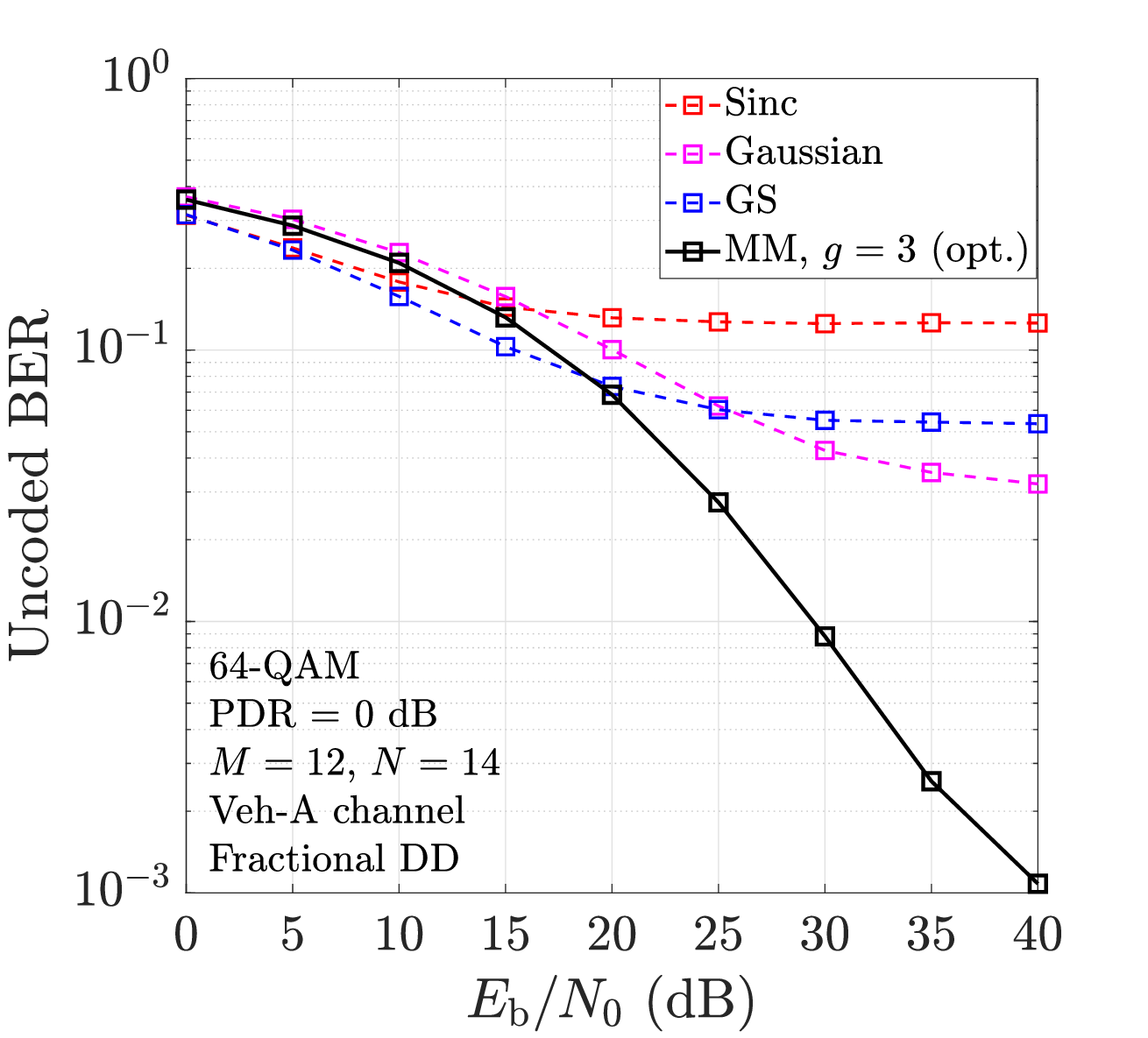}
    \subcaption{}
\end{subfigure}
\vspace{-1mm}
\caption{Uncoded BER performance comparison between 
the optimized MM filter and other benchmark filters (sinc, Gaussian, and GS): BER versus $E_\mathrm{b}/N_0$ for
(a) $4$-QAM, (b) $16$-QAM, and (c) $64$-QAM.}
\label{fig:ber_qam}
\vspace{-3mm}
\end{figure*}

{\em Uncoded BER Performance:}  
Figs.~\ref{fig:ber_qam}(a), (b), and (c) show the uncoded BER performance achieved by different filters as a function of $E_\mathrm{b}/N_0$ for 4-QAM, 16-QAM, and 64-QAM, respectively. 
It is observed that while all filters achieve roughly similar performance at low $E_\mathrm{b}/N_0$ values, the optimized MM filter outperforms other filters in the medium-to-high $E_\mathrm{b}/N_0$ values. Also, as the modulation order increases, the performance gain in favor of the optimized MM filter becomes more prominent owing to the compactness of the optimized MM pulse, which mitigates pilot-data interference leading to improved I/O relation estimation and detection performance. For example, for 64-QAM, the optimized MM filter achieves more than an order-of-magnitude improvement in BER compared to other filters at $40$~dB $E_\mathrm{b}/N_0$: $1\times 10^{-3}$ BER for optimized MM filter vs $> 3\times 10^{-2}$ BER for other filters.

\begin{figure*}[t]
\centering
\begin{subfigure}[b]{0.32\textwidth}
    \includegraphics[width=\textwidth,height=0.75\textwidth]{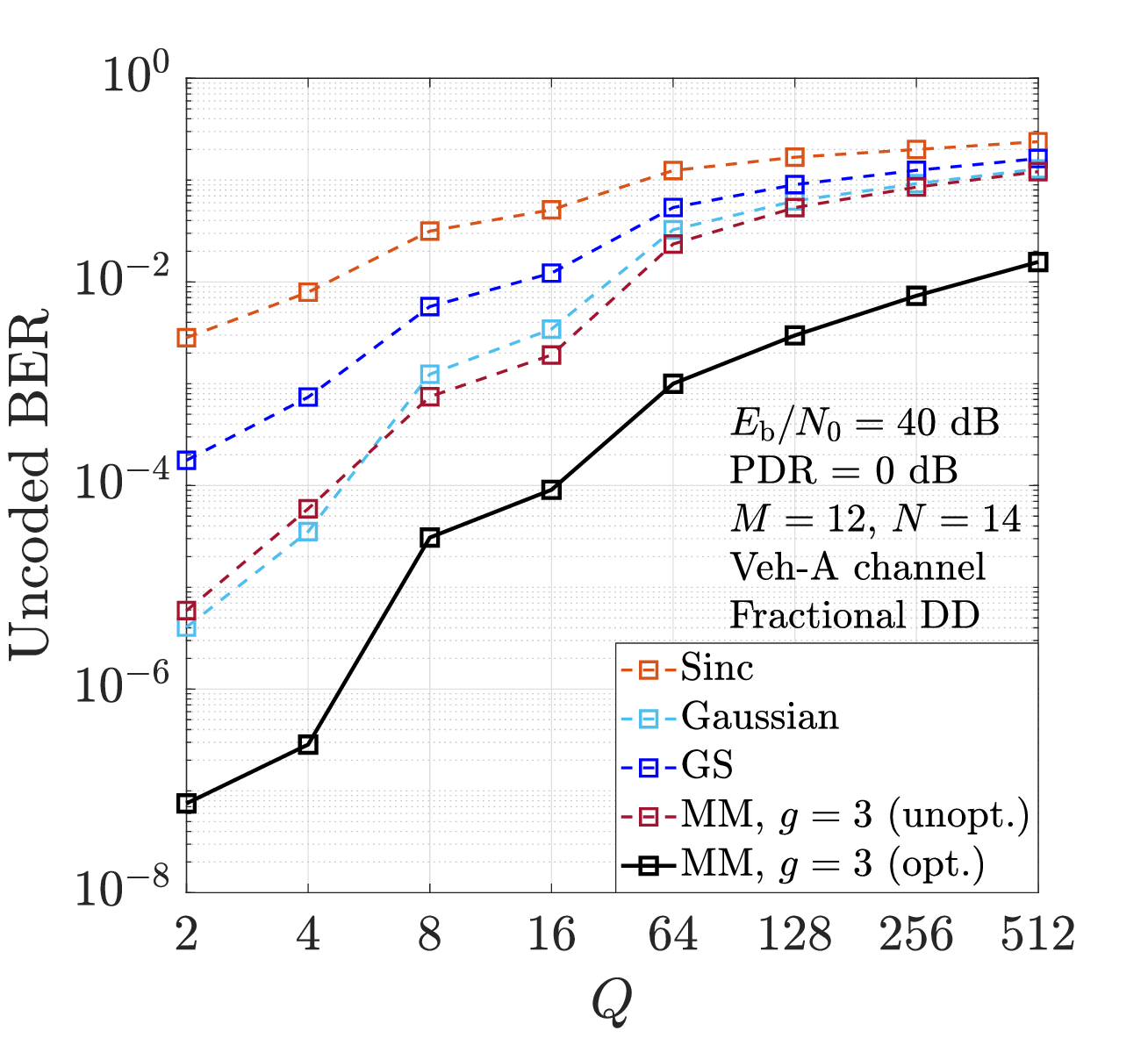}
    \subcaption{}
\end{subfigure}
\begin{subfigure}[b]{0.32\textwidth}
    \includegraphics[width=\textwidth,height=0.75\textwidth]{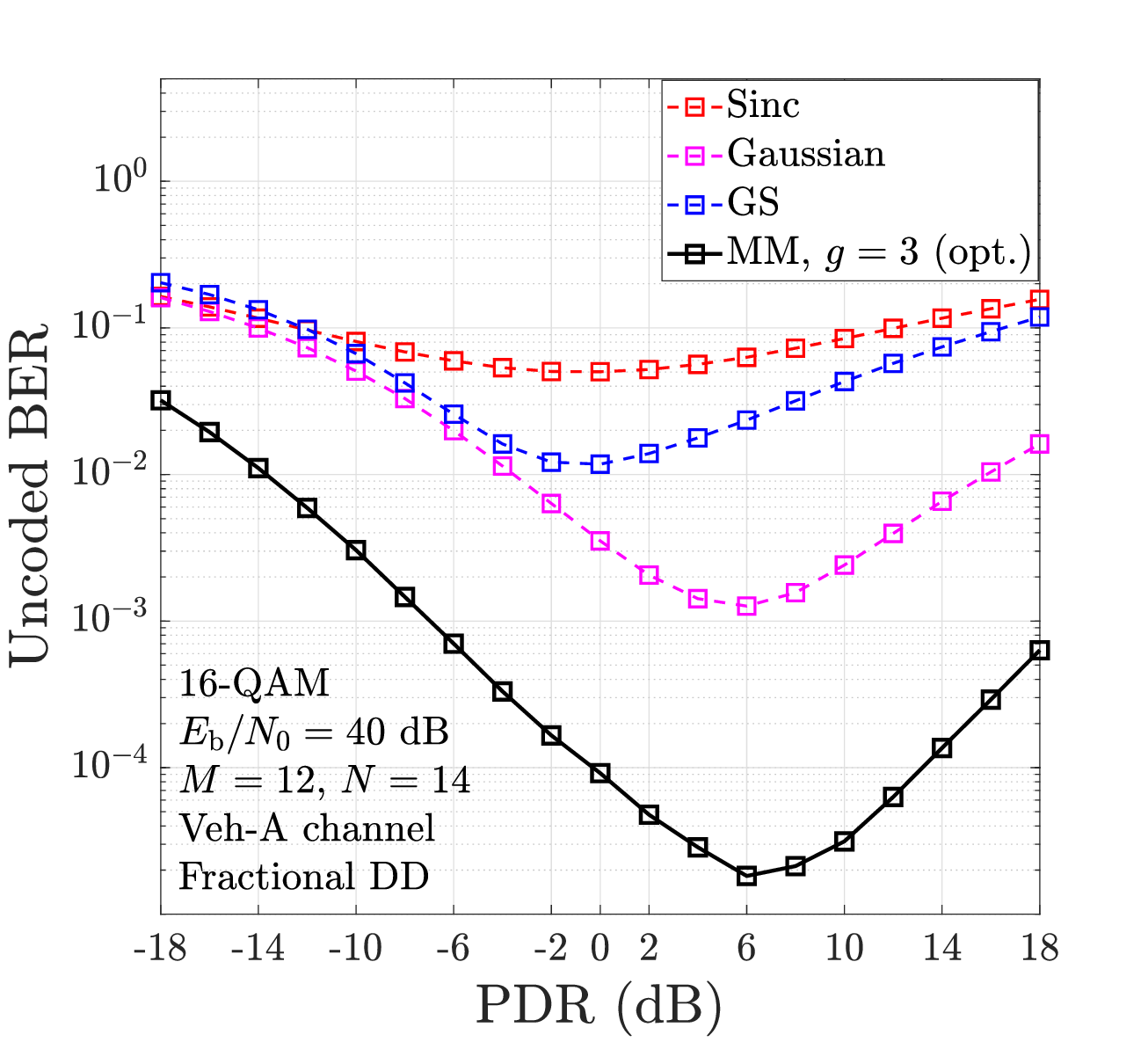}
    \subcaption{}
\end{subfigure}
\begin{subfigure}[b]{0.32\textwidth}
    \includegraphics[width=\textwidth,height=0.75\textwidth]{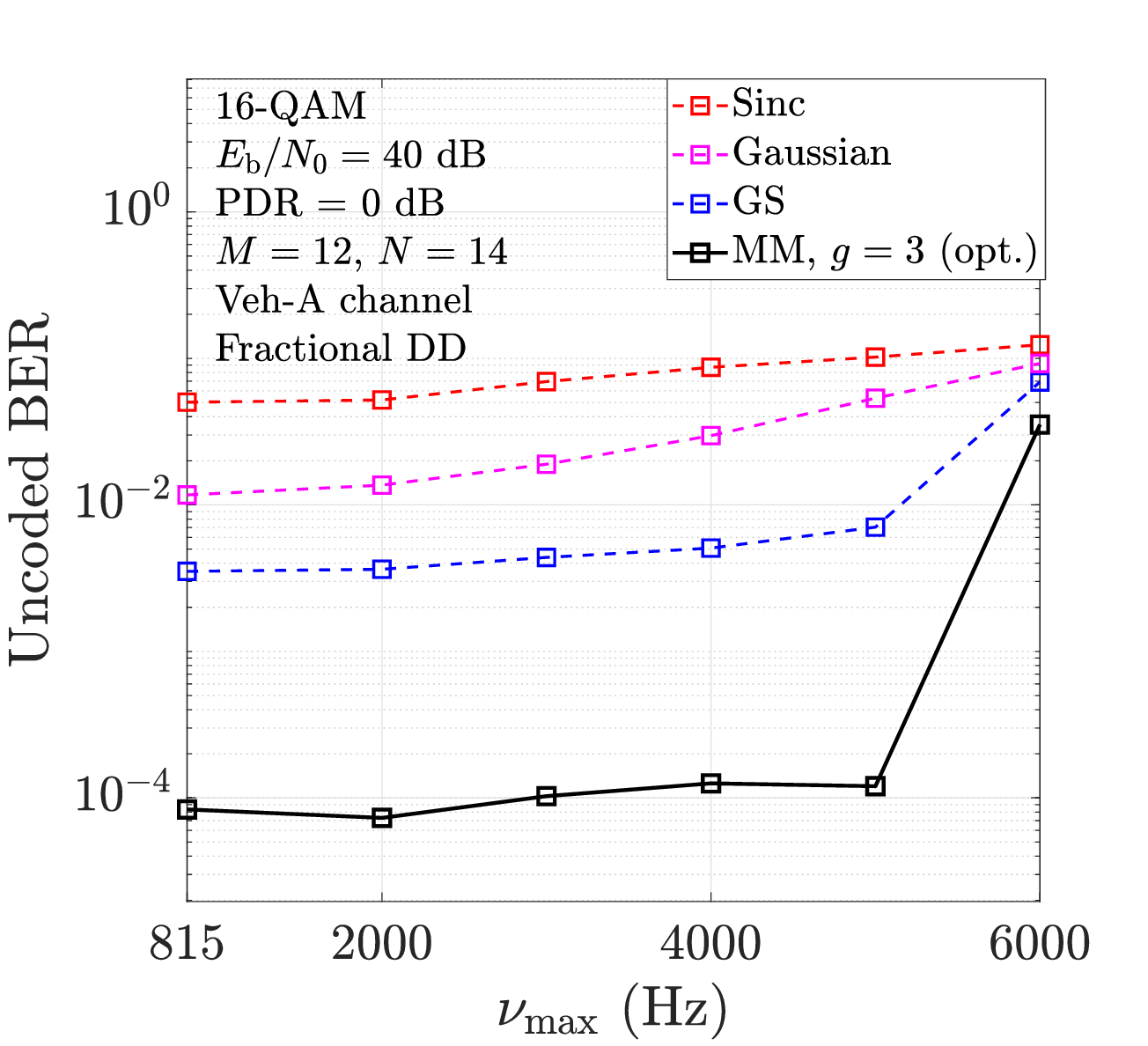}
    \subcaption{}
\end{subfigure}
\vspace{-1mm}
\caption{Uncoded BER performance comparison between the optimized MM filter and other benchmark filters (sinc, Gaussian, and GS): (a) BER vs modulation order $Q$ at $40$~dB $E_\text{b}/N_0$, (b) BER vs PDR for $16$-QAM at $40$~dB $E_\text{b}/N_0$, and (c) BER vs maximum Doppler $\nu_{\mathrm{max}}$ for $16$-QAM at $40$~dB $E_\text{b}/N_0$.}
\label{fig:ber_other}
\vspace{-3mm}
\end{figure*}

In Fig.~\ref{fig:ber_other}(a), we compare the BER performance of different filters as a function of the modulation order $(Q)$, from BPSK to $512$-QAM, at $E_\mathrm{b}/N_0=40$ dB. It can be observed that the optimized MM filter consistently achieves the lowest BER, maintaining a clear performance advantage as the modulation order increases.

In Fig.~\ref{fig:ber_other}(b), we show the BER as a function of PDR for $16$-QAM at $E_\mathrm{b}/N_0=40$~dB. The performance curve follows the expected U-shaped trend: at negative PDRs ($E_\mathrm{p} < E_\mathrm{d}$), the pilot energy is lower than the data energy, leading to poor channel estimation from a weak pilot. Conversely, at positive PDRs ($E_\mathrm{p} > E_\mathrm{d}$), data symbol power is reduced while stronger pilot-data interference causes the BER to rise. Here again, the optimized MM filter achieves significantly lower BER compared to other filters across the considered PDR range, consistent with its highly compact, sidelobe-free structure which helps to keep pilot-data interference low.

In Fig.~\ref{fig:ber_other}(c), we show the BER performance as a function of the maximum Doppler $\nu_{\max}$ for $16$-QAM at $40$ dB $E_\mathrm{b}/N_0$. Because of the Doppler resilience of Zak-OTFS, the BER curves remain relatively flat across the Doppler range up to $\nu_{\max} = 5$ kHz, with the optimized MM filter achieving BER more than an order-of-magnitude better compared to other filters.

\begin{figure}
\centering
\includegraphics[width=0.95\linewidth,height=0.75\linewidth]{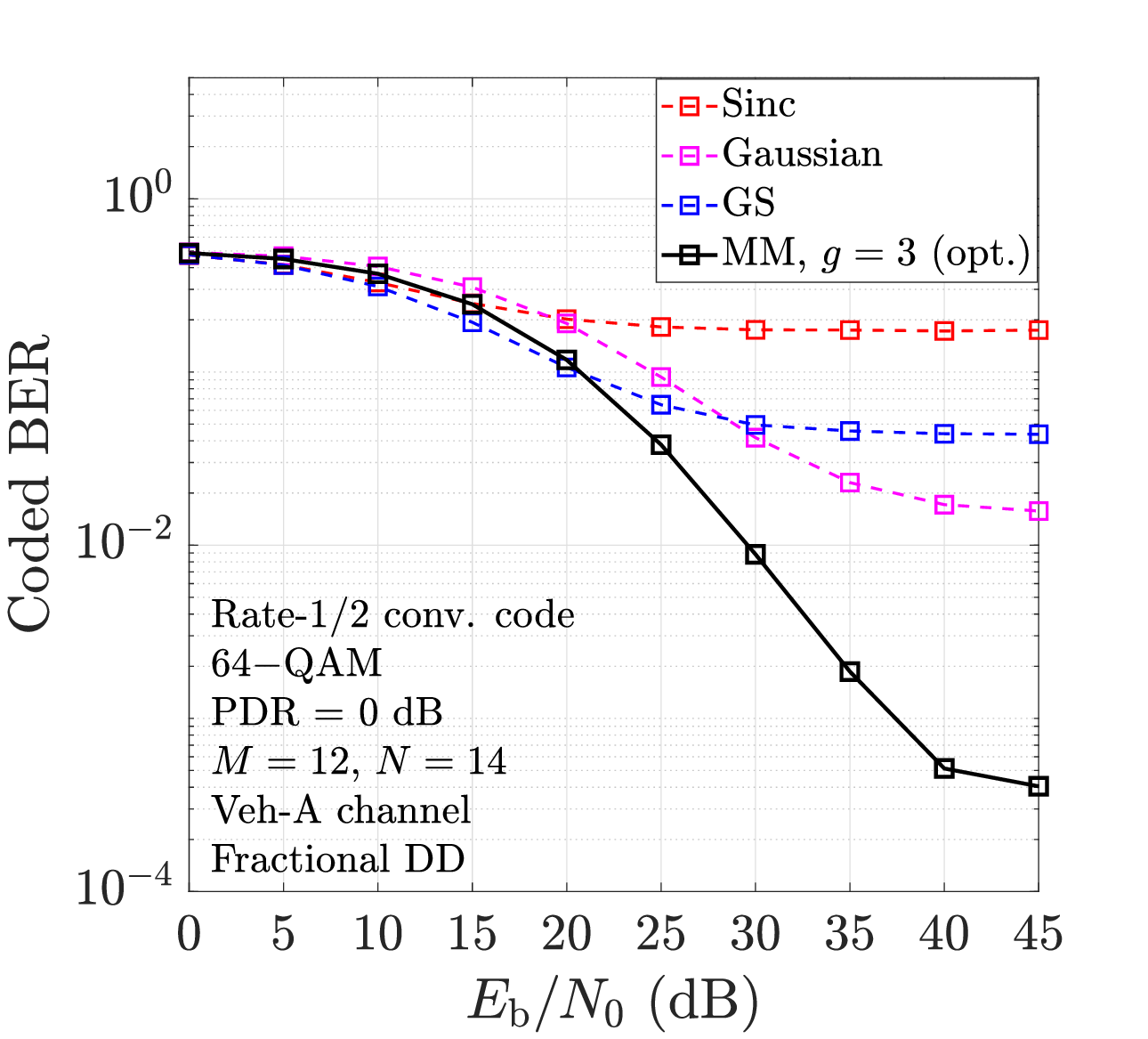}
\caption{Coded BER performance comparison between optimized MM filter and other benchmark filters for $64$-QAM, rate-1/2 convolutional code.}
\label{fig:coded}
\vspace{-4mm}
\end{figure}

{\em Coded BER Performance:} Fig.~\ref{fig:coded} compares the coded BER performance of $64$-QAM for different filters using a rate-$1/2$ convolutional code with constraint length $7$. The optimized MM filter achieves more than an order-of-magnitude improvement in coded BER compared to other filters. For example, at 45 dB $E_\mathrm{b}/N_0$, the optimized MM filter achieves a coded BER of $3\times10^{-4}$, whereas the coded BER of other filters floors at high error rates above $10^{-2}$.

\begin{figure}
\centering
\includegraphics[width=0.95\linewidth,height=0.75\linewidth]{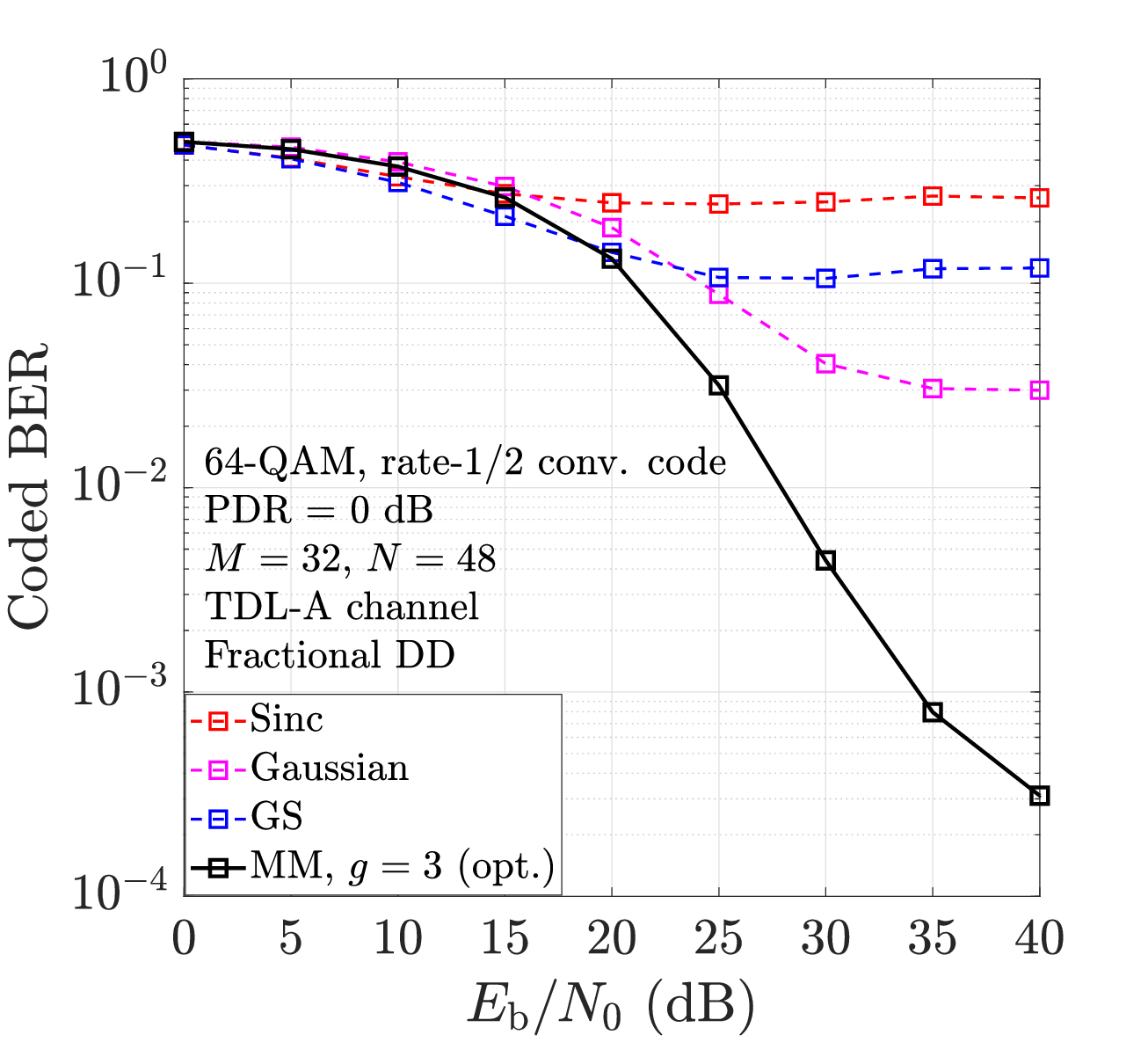}
\caption{Coded BER performance for larger frame size ($M = 32$, $N = 48$) and TDL-A channel model.}
\vspace{-2mm}
\label{fig:tdla}
\end{figure}

{\em Performance for larger frame and other channel models:}
We further assess the coded BER performance under larger frame sizes and other channel models. Fig.~\ref{fig:tdla} presents the coded BER performance of $64$-QAM under the TDL-A channel model with $P=23$ paths \cite{tdla}, $\tau_{\max}=2.89~\mu$s, and $\nu_{\max} = 815$ Hz. We consider a larger frame size with $M=32$ and $N=48$. For the considered delay and Doppler periods of $\tau_\mathrm{p}=66.67$ $\mu$s and $\nu_\mathrm{p}=15$ kHz, these $M$ and $N$ values correspond to a bandwidth of $B=480$~kHz and a frame duration of $T=3.2$~ms. The frame parameters are chosen as $p_1=0$, $p_2=1$, $g_1=0$, and $g_2=1$, with the maximum delay index given by $k_{\max}=\lceil B\tau_{\max}\rceil=2$, achieving a frame utilization of $\eta=78.125\%$. For this system configuration and channel model, we see that, at $E_\mathrm{b}/N_0=40$ dB, the optimized MM filter achieves about two orders-of-magnitude improvement in coded BER compared to the benchmark filters. 

\section{Conclusions}
\label{sec:conclusions}
Recognizing the shortfall in the performance of existing DD pulse shaping filters in Zak-OTFS for higher-order modulation, we developed a separable two-dimensional DD filter under the MM pulse design framework and proposed a coefficient optimization framework for the MM pulse family that minimizes aggregate sidelobe energy while penalizing deviation from the prototype filter. The proposed optimization improved DD domain localization while retaining the compact MM structure that resulted in significantly improved performance for higher-order QAM in Zak-OTFS. For the proposed filter, we derived closed-form expressions for the effective DD domain channel and the covariance of the receive-filtered noise. These expressions significantly reduced the simulation run times. Simulation results with embedded pilot frames and model-free I/O relation estimation for higher-order QAM showed that the optimized MM filter outperformed benchmark filters in the literature such as
sinc, Gaussian, and GS filters. Significantly improved performance (e.g., more than an order-of-magnitude improvement in coded BER) was demonstrated for different channel models (Veh-A and TDL-A channel models with fractional DDs) and frame sizes. Future work can investigate joint optimization of DD domain localization and grid-point suppression and extension to integrated sensing and communication environments. 

\appendices

\section{Proof of Theorem \ref{thm:closed_form_heff}}
\label{app:heff}
From the definition of the effective channel and the matched receive
filter,
$
h_{\mathrm{eff}}(\tau,\nu)
=
w_{\mathrm{rx}}(\tau,\nu)
*_{\sigma}
h(\tau,\nu)
*_{\sigma}
w_{\mathrm{tx}}(\tau,\nu)$.
Using $h(\tau,\nu)=\sum_{i=1}^{P}h_i\delta(\tau-\tau_i)\delta(\nu-\nu_i)$, we first obtain
\begin{equation}
(h *_{\sigma} w_{\mathrm{tx}})(\tau,\nu) = \sum_{i=1}^{P} h_i w_{\mathrm{tx}}(\tau-\tau_i,\nu-\nu_i)e^{j2\pi\nu_i(\tau-\tau_i)}.
\label{eq:wtx_h}
\end{equation}
Applying $w_{\mathrm{rx}}(\tau,\nu)$ to \eqref{eq:wtx_h} with $w_{\mathrm{rx}}(\tau,\nu) = w_\tau^{*}(-\tau)w_\nu^{*}(-\nu)e^{j2\pi\tau\nu}$ and exploiting pulse separability, the cascade separates into delay- and Doppler-domain integrals for each path as
\begin{equation}
h_{\mathrm{eff}}(\tau,\nu)
=
\sum_{i=1}^{P}
h_i
e^{j2\pi\nu_i(\tau-\tau_i)}
I_{\tau,i}(\tau)
I_{\nu,i}(\tau,\nu),
\label{eq:heff_app_cascade}
\end{equation}
where
\begin{eqnarray}
\hspace{-6mm}
I_{\tau,i}(\tau)
& \hspace{-2mm} \triangleq & \hspace{-2mm}
\int_{-\infty}^{\infty}
w_\tau^{*}(-\tau')
w_\tau(\tau-\tau_i-\tau')
e^{-j2\pi\nu_i\tau'}
\,d\tau',
\label{eq:I_tau_app}
\end{eqnarray}
\begin{eqnarray}
\hspace{-6mm}
I_{\nu,i}(\tau,\nu)
& \hspace{-2mm} \triangleq \hspace{-2mm} &
\int_{-\infty}^{\infty}
w_\nu^{*}(-\nu')
w_\nu(\nu-\nu_i-\nu')
e^{j2\pi\tau\nu'}
\,d\nu'.
\label{eq:I_nu_app}
\end{eqnarray}
We first evaluate the delay-domain term in \eqref{eq:I_tau_app}.
Using \eqref{p_tau} and the identity $\cos(x+\pi\ell)=(-1)^\ell\cos(x)$, the delay-domain shaping function can be written as
\begin{align}
p_\tau(\tau)
&=
c_0+
2\sum_{\ell=1}^{g-1}
(-1)^\ell c_\ell
\cos\left(
\frac{2\pi\ell B}{\zeta_{\tau}}\tau
\right)
\nonumber\\
&=
\sum_{m=-(g-1)}^{g-1}
a_m
e^{j\frac{2\pi mB}{\zeta_{\tau}}\tau},
\label{eq:ptau_app}
\end{align}
where the coefficients $a_m$ are those defined in Theorem~\ref{thm:closed_form_heff}. Consequently,
\begin{equation}
w_\tau(\tau)
=
\kappa_{\tau}\sqrt{B}
\sum_{m=-(g-1)}^{g-1}
a_m
e^{j\frac{2\pi mB}{\zeta_{\tau}}\tau}
\operatorname{rect}
\left(
\frac{B\tau}{\zeta_{\tau}}
\right).
\label{eq:wtau_app}
\end{equation}
Substituting \eqref{eq:wtau_app} into \eqref{eq:I_tau_app}, noting that conjugation and time-reversal in $w_\tau^{*}(-\tau')$ yield the positive-exponent phase $e^{j2\pi mB\tau'/\zeta_{\tau}}$, and interchanging the finite summations and integration gives
\begin{align}
I_{\tau,i}(\tau)
&=
\kappa_{\tau}^2 B
\sum_{m,n=-(g-1)}^{g-1}
a^*_m
a_n
e^{j\frac{2\pi nB}{\zeta_{\tau}}(\tau-\tau_i)}
J_{mn}^{(i)}(\tau),
\label{eq:I_tau_sum}
\end{align}
where
\begin{eqnarray}
J_{mn}^{(i)}(\tau)
& \hspace{-2mm} \triangleq & \hspace{-2mm}
\int_{-\infty}^{\infty}
e^{j2\pi\beta_{mn}^{(i)}\tau'}
\operatorname{rect}\left(
\frac{B\tau'}{\zeta_{\tau}}
\right) \nonumber \\  
& \hspace{-2mm} & \hspace{-2mm} \times
\operatorname{rect}\left(
\frac{B(\tau-\tau_i-\tau')}{\zeta_{\tau}}
\right)
\!d\tau'.
\label{eq:J_mn_app}
\end{eqnarray}
Let $L_\tau \triangleq \zeta_{\tau}/(2B)$. The two rectangular windows in \eqref{eq:J_mn_app} are supported on $[-L_\tau,L_\tau]$ and $[\tau-\tau_i-L_\tau,\, \tau-\tau_i+L_\tau]$, respectively, which overlap if and only if $|\tau-\tau_i| \leq \zeta_{\tau}/B$. Their intersection is centered at $C_{\tau,i}(\tau) = (\tau-\tau_i)/2$ with overlap length $\Lambda_{\tau,i}(\tau) = [\zeta_{\tau}/B - |\tau-\tau_i|]^+$ as defined in \eqref{eq:Lambda_tau} (which equals zero when there is no overlap). Therefore,
\begin{align}
J_{mn}^{(i)}(\tau)
&=
\int_{C_{\tau,i}(\tau)-\frac{\Lambda_{\tau,i}(\tau)}{2}}
^{C_{\tau,i}(\tau)+\frac{\Lambda_{\tau,i}(\tau)}{2}}
e^{j2\pi\beta_{mn}^{(i)}\tau'}
\,d\tau'
\nonumber\\
&=
\Lambda_{\tau,i}(\tau)
e^{j\pi\beta_{mn}^{(i)}(\tau-\tau_i)}
\operatorname{sinc}\left(
\beta_{mn}^{(i)}
\Lambda_{\tau,i}(\tau)
\right).
\label{eq:J_mn_closed}
\end{align}
Substituting \eqref{eq:J_mn_closed} into \eqref{eq:I_tau_sum} and
using the definition of $\beta_{mn}^{(i)}$ yields
\begin{align}
I_{\tau,i}(\tau)
&=
\kappa_{\tau}^2 B
\Lambda_{\tau,i}(\tau)
\sum_{m,n=-(g-1)}^{g-1}
a^*_m
a_n
\nonumber\\
&\quad\times
\exp\left\{
j\pi
\left(
\frac{(m+n)B}{\zeta_{\tau}}
-\nu_i
\right)
(\tau-\tau_i)
\right\}
\nonumber\\
&\quad\times
\operatorname{sinc}\left(
\beta_{mn}^{(i)}
\Lambda_{\tau,i}(\tau)
\right).
\label{eq:I_tau_closed}
\end{align}
The Doppler-domain term in \eqref{eq:I_nu_app} is evaluated
analogously. From \eqref{p_nu}, the Doppler-domain pulse admits the
finite exponential representation
\begin{equation}
w_\nu(\nu)
=
\kappa_{\nu}\sqrt{T}
\sum_{r=-(g-1)}^{g-1}
a_r
e^{j\frac{2\pi rT}{\zeta_{\nu}}\nu}
\operatorname{rect}
\left(
\frac{T\nu}{\zeta_{\nu}}
\right).
\label{eq:wnu_app}
\end{equation}
Substitution of \eqref{eq:wnu_app} into \eqref{eq:I_nu_app} gives
\begin{align}
I_{\nu,i}(\tau,\nu)
&=
\kappa_{\nu}^2 T
\!\!\!\!\sum_{r,s=-(g-1)}^{g-1}
a^*_r
a_s
e^{j\frac{2\pi sT}{\zeta_{\nu}}(\nu-\nu_i)}
\widetilde J_{rs}^{(i)}(\tau,\nu),
\label{eq:I_nu_sum}
\end{align}
where
\begin{eqnarray}
\widetilde J_{rs}^{(i)}(\tau,\nu)
& \hspace{-2mm} \triangleq & \hspace{-2mm}
\!\!\!\int_{-\infty}^{\infty}
\!\!\!e^{j2\pi\gamma_{rs}(\tau)\nu'}
\operatorname{rect}\left(
\frac{T\nu'}{\zeta_{\nu}}
\right) \nonumber \\ 
& \hspace{-2mm} & \hspace{-2mm} \times
\operatorname{rect}\left(
\frac{T(\nu-\nu_i-\nu')}{\zeta_{\nu}}
\right)
\!d\nu'.
\label{eq:Jtilde_def}
\end{eqnarray}
Following the same rectangular-overlap argument as in the delay-domain case, the two windows in~\eqref{eq:Jtilde_def} have overlap length $\Lambda_{\nu,i}(\nu)$ as defined in~\eqref{eq:Lambda_nu}.
Hence, using the same rectangular-overlap integral as in
\eqref{eq:J_mn_closed},
\begin{equation}
\hspace{-2.5mm}
\widetilde J_{rs}^{(i)}(\tau,\nu)
 = 
\Lambda_{\nu,i}(\nu)
e^{j\pi\gamma_{rs}(\tau)(\nu-\nu_i)}
\operatorname{sinc}\left(
\gamma_{rs}(\tau)\Lambda_{\nu,i}(\nu)
\right). \hspace{-1.5mm}
\end{equation}
Substitution into \eqref{eq:I_nu_sum} and collection of the
exponential terms yield
\vspace{-2mm}
\begin{align}
I_{\nu,i}(\tau,\nu)
&=
\kappa_{\nu}^2 T
\Lambda_{\nu,i}(\nu)
\sum_{r,s=-(g-1)}^{g-1}
a^*_r
a_s
\nonumber\\
&\quad\times
\exp\left\{
j\pi
\left(
\tau+\frac{(r+s)T}{\zeta_{\nu}}
\right)
(\nu-\nu_i)
\right\}
\nonumber\\
&\quad\times
\operatorname{sinc}\left(
\gamma_{rs}(\tau)\Lambda_{\nu,i}(\nu)
\right).
\label{eq:I_nu_closed}
\end{align}
Substituting \eqref{eq:I_tau_closed} and \eqref{eq:I_nu_closed} into \eqref{eq:heff_app_cascade}, and combining the exponential terms, the resulting phase is
\begin{align}
\Phi_{mnrs}^{(i)}(\tau,\nu)
&= 2\pi\nu_i(\tau-\tau_i)
+ \pi\left(
\frac{(m+n)B}{\zeta_{\tau}}-\nu_i
\right)(\tau-\tau_i)
\nonumber\\
&\quad+
\pi\left(
\tau+\frac{(r+s)T}{\zeta_{\nu}}
\right)(\nu-\nu_i)
\nonumber\\
&=
\pi\left(
\nu_i+\frac{(m+n)B}{\zeta_{\tau}}
\right)(\tau-\tau_i)
\nonumber\\
&\quad+
\pi\left(
\tau+\frac{(r+s)T}{\zeta_{\nu}}
\right)(\nu-\nu_i),
\end{align}
which is precisely the phase term in \eqref{eq:Phi_mnrs}.
Hence, \eqref{eq:heff_app_cascade} reduces to \eqref{eq:heff_cf},
proving Theorem~\ref{thm:closed_form_heff}.

\section{Proof of Theorem \ref{thm:noise_cov}}
\label{app:noise_cov}

The Zak transform of the time-domain AWGN is
\begin{equation}
n_{\mathrm{dd}}(\tau,\nu)
=
\sqrt{\tau_{\mathrm{p}}}
\sum_{q\in\mathbb Z}
n_{\rm td}(\tau+q\tau_{\mathrm{p}})e^{-j2\pi\nu q\tau_{\mathrm{p}}}.
\end{equation}
The receive-filtered DD domain noise is
\begin{equation}
n_{\mathrm{dd}}^{w_{\mathrm{rx}}}(\tau,\nu)
=
w_{\mathrm{rx}}(\tau,\nu)*_{\sigma}
n_{\mathrm{dd}}(\tau,\nu).
\end{equation}
Using the matched receive filter and the definition of twisted convolution,
\begin{align}
n_{\mathrm{dd}}^{w_{\mathrm{rx}}}(\tau,\nu)
&=
\iint
w_{\tau}^{*}(-\tau')
w_{\nu}^{*}(-\nu')
e^{j2\pi\tau'\nu'}
\nonumber\\
&\quad\times
n_{\mathrm{dd}}(\tau-\tau',\nu-\nu')
e^{j2\pi\nu'(\tau-\tau')}
d\tau'd\nu'.
\end{align}
Since $\tau'\nu'+\nu'(\tau-\tau')=\nu'\tau,$ the phase terms reduce to $e^{j2\pi\nu'\tau}$. Substituting the Zak-domain noise expression and separating the delay- and Doppler-domain terms gives
\begin{equation}
n_{\mathrm{dd}}^{w_{\mathrm{rx}}}(\tau,\nu)
=
\sqrt{\tau_{\mathrm{p}}}
\sum_{q\in\mathbb Z}
e^{-j2\pi\nu q\tau_{\mathrm{p}}}
\mathcal{G}_{\nu}(\tau+q\tau_{\mathrm{p}})
v(\tau+q\tau_{\mathrm{p}}),
\label{eq:app_noise_cont}
\end{equation}
where
\begin{equation}
v(x)
=
\int_{-\infty}^{\infty}
w_{\tau}^{*}(-u)n_{\rm td}(x-u)\,du,
\label{eq:app_v}
\end{equation}
and
\begin{equation}
\mathcal{G}_{\nu}(x)
=
\int_{-\infty}^{\infty}
w_{\nu}^{*}(-\nu')
e^{j2\pi\nu'x}\,d\nu'.
\label{eq:app_Gnu_def}
\end{equation}
Sampling \eqref{eq:app_noise_cont} at
$(\tau,\nu)=(k\Delta\tau,l\Delta\nu)$ and using
$\tau_{\mathrm{p}}\nu_{\mathrm{p}}=1$ yields
\begin{equation}
n_{\mathrm{dd}}[k,l]
=
\sqrt{\tau_{\mathrm{p}}}
\sum_{q\in\mathbb Z}
e^{-j2\pi lq/N}
\mathcal{G}_{\nu}(x_{k,q})
v(x_{k,q}),
\label{eq:app_noise_sampled}
\end{equation}
where $x_{k,q}=q\tau_{\mathrm{p}}+k\Delta\tau$. Hence, defining $x_i=q_i\tau_{\mathrm{p}}+k_i\Delta\tau$ and 
$\Delta_x=x_2-x_1$,
the noise covariance becomes
\begin{align}
&\mathbb{E}\!\left[
n_{\mathrm{dd}}[k_1,l_1]
n_{\mathrm{dd}}^{*}[k_2,l_2]
\right]
\nonumber\\
&=
\tau_{\mathrm{p}}
\sum_{q_1,q_2\in\mathbb Z}
\hspace{-2mm}
e^{-j2\pi(q_1l_1-q_2l_2)/N}
\mathcal{G}_{\nu}(x_1)
\mathcal{G}_{\nu}^{*}(x_2)
\mathbb{E}\!\left[v(x_1)v^{*}(x_2)\right].
\label{eq:app_noise_cov_intermediate}
\end{align}
Using the AWGN autocorrelation
$\mathbb{E}\!\left[n_{\rm td }(t_1)n_{\rm td}^{*}(t_2)\right]
=
N_0\delta(t_1-t_2)$,
it follows that
\begin{eqnarray}
\hspace{-6mm}
\mathbb{E}\!\left[v(x_1)v^{*}(x_2)\right]
& \hspace{-2mm} = & \hspace{-2mm}
N_0
\iint
w_{\tau}^{*}(-u_1)w_{\tau}(-u_2)
\nonumber \\
& \hspace{-2mm} & \hspace{-2mm} \times
\, \delta(u_2-u_1+x_1-x_2)
\,du_1du_2
\nonumber \\
& \hspace{-2mm} = & \hspace{-2mm}
N_0
\int_{-\infty}^{\infty}
w_{\tau}^{*}(-u)
w_{\tau}(-(u+\Delta_x))\,du.
\label{eq:app_v_cov}
\end{eqnarray}
Thus, 
$\mathbb{E}\!\left[v(x_1)v^{*}(x_2)\right]
=
N_0 C_{\tau}(\Delta_x)$,
where
\begin{equation}
C_{\tau}(\Delta_x)
\triangleq
\int_{-\infty}^{\infty}
w_{\tau}^{*}(-u)
w_{\tau}(-(u+\Delta_x))\,du.
\label{eq:app_Ctau_def}
\end{equation}
Substituting $\mathbb{E}\!\left[v(x_1)v^{*}(x_2)\right]$ into
\eqref{eq:app_noise_cov_intermediate} yields
\eqref{eq:noise_cov}. We next derive the closed-form expressions for $\mathcal{G}_{\nu}(x)$ and $C_{\tau}(\Delta_x)$.
To obtain $\mathcal{G}_{\nu}(x)$, substituting the real-valued, even Doppler-domain MM pulse into \eqref{eq:app_Gnu_def} gives
\begin{align}
\mathcal{G}_{\nu}(x)
&=
\kappa_{\nu}\sqrt{T}
\int_{-\frac{\zeta_{\nu}}{2T}}^{\frac{\zeta_{\nu}}{2T}}
p_{\nu}(-\nu')e^{j2\pi\nu'x}\,d\nu'.
\end{align}
Using the symmetry $a_{-z}=a_z$ from \eqref{eq:a_q}, the finite exponential expansion of $p_{\nu}(-\nu')$ is given by
\begin{equation}
p_{\nu}(-\nu')
=
\sum_{z=-(g-1)}^{g-1}
a_z e^{j\frac{2\pi zT}{\zeta_{\nu}}\nu'},
\end{equation}
which yields
\begin{align}
\mathcal{G}_{\nu}(x)
&=
\kappa_{\nu}\sqrt{T}
\sum_{z=-(g-1)}^{g-1}a_z
\int_{-\frac{\zeta_{\nu}}{2T}}^{\frac{\zeta_{\nu}}{2T}}
e^{j2\pi\left(x+\frac{zT}{\zeta_{\nu}}\right)\nu'}d\nu'
\nonumber\\
&=
\kappa_{\nu}\frac{\zeta_{\nu}}{\sqrt{T}}
\sum_{z=-(g-1)}^{g-1}
a_z
\operatorname{sinc}\!\left(
\frac{\zeta_{\nu}x}{T}+z
\right).
\label{eq:app_Gnu_closed}
\end{align}
It remains to evaluate $C_{\tau}(\Delta_x)$. Substituting the delay-domain MM pulse into its definition gives
\begin{align}
C_{\tau}(\Delta_x)
&=
\kappa_{\tau}^{2}B
\int_{-\infty}^{\infty}
p^*_{\tau}(-u)
p_{\tau}(-(u+\Delta_x))
\nonumber\\
&\quad\times
\operatorname{rect}\!\left(\frac{Bu}{\zeta_{\tau}}\right)
\operatorname{rect}\!\left(\frac{B(u+\Delta_x)}{\zeta_{\tau}}\right)
\,du.
\label{eq:c_rect}
\end{align}
The two rectangular windows in~\eqref{eq:c_rect} overlap only when $|\Delta_x|<\zeta_{\tau}/B$. In this case, their intersection is the interval $[u_1,u_2]$, where
\begin{equation}
u_1=
\max\!\Big(
-\frac{\zeta_{\tau}}{2B},
-\frac{\zeta_{\tau}}{2B}-\Delta_x
\Big),
\,\,
u_2=
\min\!\big(
\frac{\zeta_{\tau}}{2B},
\frac{\zeta_{\tau}}{2B}-\Delta_x
\big).
\end{equation}
Consequently, the length of the overlap interval is
\begin{equation}
u_2-u_1
=
\Gamma_{\tau}(\Delta_x)
=
\left[
\frac{\zeta_{\tau}}{B}-|\Delta_x|
\right]^{+}.
\end{equation}
For $|\Delta_x|\geq\zeta_{\tau}/B$, the overlap has zero measure and hence $C_{\tau}(\Delta_x)=0$. For $|\Delta_x|<\zeta_{\tau}/B$, substituting the finite exponential expansions of $p_{\tau}^{*}(-u)$ and $p_{\tau}(-(u+\Delta_x))$ with coefficients $a_m^{*}$ and $a_n$ from \eqref{eq:a_q} gives
\begin{align}
C_{\tau}(\Delta_x)
&=
\kappa_{\tau}^{2}B
\sum_{m,n=-(g-1)}^{g-1}
a_m^{*}a_n
\nonumber\\
&\quad\times
\int_{u_1}^{u_2}
e^{j\frac{2\pi B}{\zeta_{\tau}}(m-n)u}
e^{-j\frac{2\pi nB}{\zeta_{\tau}}\Delta_x}
\,du.
\label{eq:app_Ctau_integral}
\end{align}
Since $[u_1,u_2]$ has midpoint $(u_1+u_2)/2 = -\Delta_x/2$ and length $u_2-u_1=\Gamma_{\tau}(\Delta_x)$, the integration limits are $u_1 = -\Delta_x/2 - \Gamma_{\tau}(\Delta_x)/2$ and $u_2 = -\Delta_x/2 + \Gamma_{\tau}(\Delta_x)/2$. Using these limits and $\operatorname{sinc}(x)\triangleq\sin(\pi x)/(\pi x)$, the integral evaluates to
\begin{eqnarray}
\hspace{-5mm}
\int_{u_1}^{u_2}
e^{j\frac{2\pi B}{\zeta_{\tau}}(m-n)u}
\,du & & \nonumber\\
& \hspace{-62mm} = & \hspace{-32mm}
\Gamma_{\tau}(\Delta_x)
e^{-j\frac{\pi B}{\zeta_{\tau}}(m-n)\Delta_x}
\operatorname{sinc}\!\left(
\frac{B}{\zeta_{\tau}}(m-n)
\Gamma_{\tau}(\Delta_x)
\right).
\end{eqnarray}
Therefore,
\begin{align}
C_{\tau}(\Delta_x)
&=
\kappa_{\tau}^{2}B\,
\Gamma_{\tau}(\Delta_x)
\sum_{m,n=-(g-1)}^{g-1}
a_m^{*}a_n
e^{-j\frac{\pi B}{\zeta_{\tau}}(m+n)\Delta_x}
\nonumber\\
&\quad\times
\operatorname{sinc}\!\left(
\frac{B}{\zeta_{\tau}}(m-n)
\Gamma_{\tau}(\Delta_x)
\right),
\label{eq:app_Ctau_closed}
\end{align}
which gives \eqref{eq:noise_Ctau}. This completes the proof.


\end{document}